\documentclass[a4paper,twocolumn,11pt,accepted=2023-09-26]{quantumarticle}

\pdfoutput=1
\usepackage[utf8]{inputenc}
\usepackage[english]{babel}
\usepackage[T1]{fontenc}
\usepackage{amsmath}
\usepackage{hyperref}
\usepackage{tikz}
\usepackage{lipsum}
\usepackage{amsfonts,amsmath,amssymb,stmaryrd}
\usepackage{url}
\usepackage{bm}
\usepackage{ulem}
\usepackage[numbers]{natbib}

\renewcommand{\l}{\left(}
\renewcommand{\r}{\right)}

\newcommand{\ket}[1]{|#1\rangle}

\renewcommand{\H}{\hat{\mathcal{H}}}

\renewcommand{\a}{\hat{a}}

\newcommand{\ad}{\hat{a}^\dagger}

\DeclareMathOperator*{\argmax}{argmax}

\newcommand{\hc}{\text{h.c.}}

\renewcommand{\sf}{\text{MIX}}

\newcommand{\Zt}{$\mathbb{Z}_2$ }

\usepackage{cancel,ifthen}
\newcommand{\cmnt}[2][NoInPuT]{\ifthenelse{\equal{#1}{NoInPuT}}{}{{\color{red}\sout{#1}}} {\color{blue} #2}}

\begin{document}

\title{Adaptive Quantum State Tomography with Active Learning}

\author{Hannah Lange}
\affiliation{Department of Physics and Arnold Sommerfeld Center for Theoretical Physics (ASC), Ludwig-Maximilians-Universit\"at M\"unchen, Theresienstr. 37, M\"unchen D-80333, Germany}
\affiliation{Munich Center for Quantum Science and Technology (MCQST), Schellingstr. 4, D-80799 M\"unchen, Germany}

\author{Matja\v{z} Kebri\v{c}}
\affiliation{Department of Physics and Arnold Sommerfeld Center for Theoretical Physics (ASC), Ludwig-Maximilians-Universit\"at M\"unchen, Theresienstr. 37, M\"unchen D-80333, Germany}
\affiliation{Munich Center for Quantum Science and Technology (MCQST), Schellingstr. 4, D-80799 M\"unchen, Germany}

\author{Maximilian Buser}
\affiliation{Department of Physics and Arnold Sommerfeld Center for Theoretical Physics (ASC), Ludwig-Maximilians-Universit\"at M\"unchen, Theresienstr. 37, M\"unchen D-80333, Germany}
\affiliation{Munich Center for Quantum Science and Technology (MCQST), Schellingstr. 4, D-80799 M\"unchen, Germany}

\author{Ulrich Schollw\"ock}
\affiliation{Department of Physics and Arnold Sommerfeld Center for Theoretical Physics (ASC), Ludwig-Maximilians-Universit\"at M\"unchen, Theresienstr. 37, M\"unchen D-80333, Germany}
\affiliation{Munich Center for Quantum Science and Technology (MCQST), Schellingstr. 4, D-80799 M\"unchen, Germany}

\author{Fabian Grusdt}
\affiliation{Department of Physics and Arnold Sommerfeld Center for Theoretical Physics (ASC), Ludwig-Maximilians-Universit\"at M\"unchen, Theresienstr. 37, M\"unchen D-80333, Germany}
\affiliation{Munich Center for Quantum Science and Technology (MCQST), Schellingstr. 4, D-80799 M\"unchen, Germany}

\author{Annabelle Bohrdt}
\affiliation{ITAMP, Harvard-Smithsonian Center for Astrophysics, Cambridge, MA 02138, USA}
\affiliation{Department of Physics, Harvard University, Cambridge, MA 02138, USA}
\maketitle

\begin{abstract}
Recently, tremendous progress has been made in the field of quantum science and technologies: different platforms for quantum simulation as well as quantum computing, ranging from superconducting qubits to neutral atoms, are starting to reach unprecedentedly large systems. In order to benchmark these systems and gain physical insights, the need for efficient tools to characterize quantum states arises. 
The exponential growth of the Hilbert space with system size renders a full reconstruction of the quantum state prohibitively demanding in terms of the number of necessary measurements.
Here we propose and implement an efficient  scheme for quantum state tomography using active learning. Based on a few initial measurements, the active learning protocol proposes the next measurement basis, designed to yield the maximum information gain.
We apply the active learning quantum state tomography scheme to reconstruct different multi-qubit states with varying degree of entanglement as well as to ground states of the XXZ model in 1D and a kinetically constrained spin chain. In all cases, we obtain a significantly improved reconstruction as compared to a reconstruction based on the exact same number of measurements and measurement configurations, but with randomly chosen basis configurations.
Our scheme is highly relevant to gain physical insights in quantum many-body systems as well as for benchmarking and characterizing quantum devices, e.g. for quantum simulation, and paves the way for scalable adaptive protocols to probe, prepare, and manipulate quantum systems.
\end{abstract}

Since the turn of this century, the characterization of quantum states is of high importance: It is needed for assessing the performance of quantum algorithms for quantum computers, certifying the quality of experimental quantum hardware, and enables the investigation of complex quantum systems \cite{Nimbe2021, Bloch2012,Preskill2018,Altman2021}. Hence, Quantum State Tomography (QST), which is the process of reconstructing quantum states from measurements, is of high interest for both experimental and theoretical research in the field of quantum physics and quantum computing. The very nature of quantum states makes the state reconstruction a complicated task and available techniques that use a full parameterization of the wave function / density operator, e.g. linear inversion or the maximum likelihood method \cite{Haeffner2005, Hradil1997}, undergo an exponential scaling of the number of samples needed for the reconstruction with the size of the quantum system to achieve a desired threshold accuracy, which is due to the exponential growth of the Hilbert space with system size. Therefore the application of conventional tomography methods is mostly limited to only a few qubits \cite{Abhijith2020}. Other works circumvent this problem by exploiting certain known state properties, e.g. MPS tomography \cite{Baumgratz2013,Lanyon2017}.

A relatively new approach is neural network quantum state tomography. As shown by Carleo and Troyer \cite{Carleo2017}, neural network representations of quantum many body wave functions are possible in many cases of interest, which can reduce the complexity to a tractable number of parameters. When training the neural quantum states on measurement data, the neural network weights capture specific structures in the input data and generalize to the full state representation \cite{Carleo2017,Torlai2018, Melko2019,Carrasquilla2019, Carrasquilla2021}. Furthermore, neural network representations can overcome limitations of non machine learning approaches, e.g. restricted Boltzmann machines can encode quantum states with an entanglement structure that goes beyond the area law due to their non-local connections between hidden and visible nodes that will be presented later in the text \cite{Chen2018,Melko2019}, in contrast to matrix product state (MPS) representations of quantum states that obey the area law of entanglement \cite{Schollwoeck2011}.  It has been shown that neural network based approaches allow the reconstruction of highly entangled states with more than a hundred qubits from a limited number of measurements, for example by using a restricted Boltzmann machine (RBM) \cite{Torlai2018,Melko2019}. Other neural network representations of quantum states include among others recurrent neural networks \cite{Morawetz2021,HibatAllah2020,Czischek2022}, variational autoencoders \cite{Rocchetto2018}, convolutional neural networks \cite{Schmale2021}, generative adversarial networks \cite{Ahmed2021} and transformer architectures \cite{Cha2021,Zhang2023}. Furthermore, machine learning based tomography can allow accessing observables which cannot be directly inferred from an experiment itself \cite{Torlai2019}.

\begin{figure}[t]
	\centering
  \includegraphics[width=0.495\textwidth]{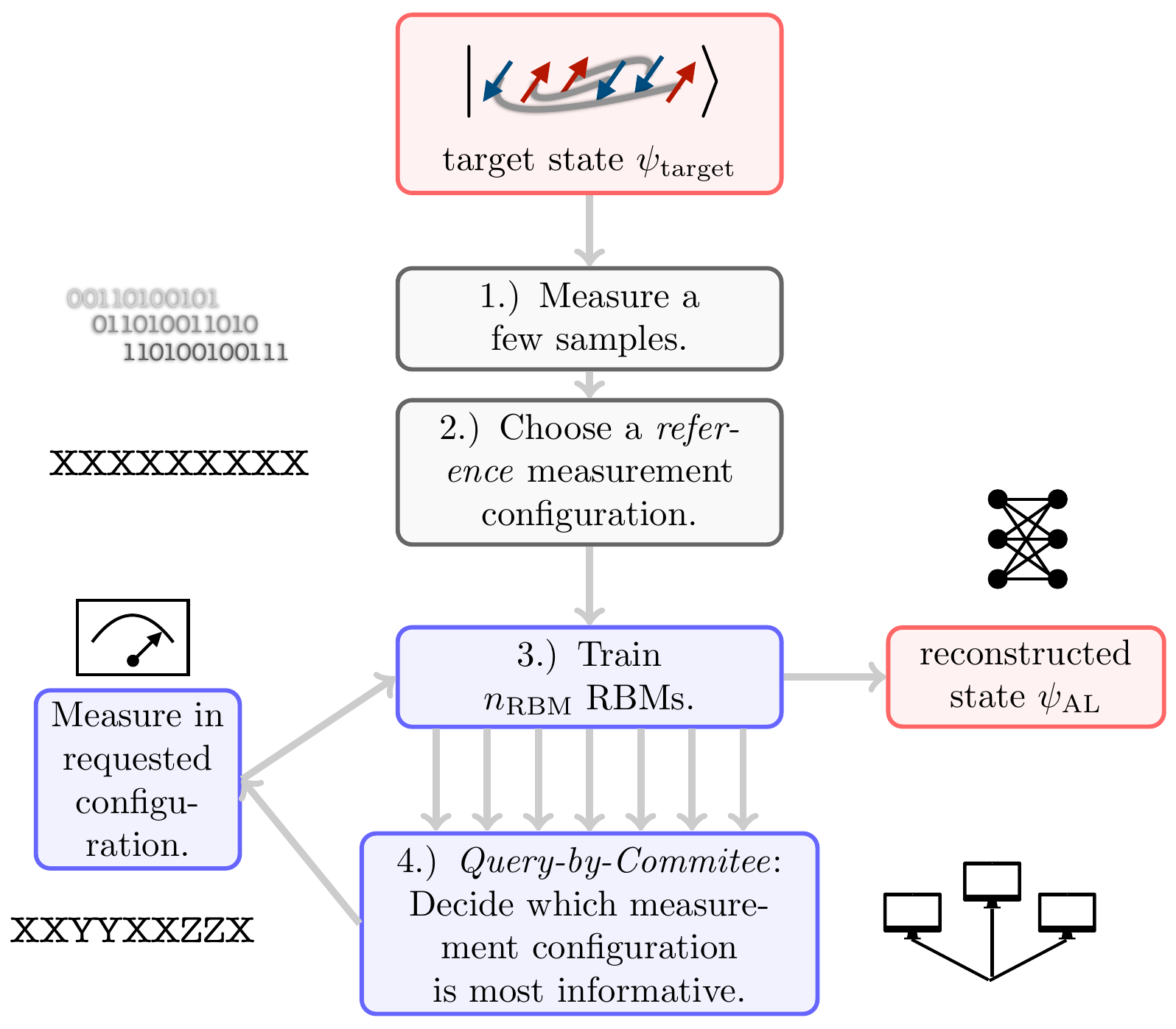}
	\caption[Active learning procedure]{Quantum state tomography (QST) by active learning. We show our active learning cycle: In the first stage a reference measurement basis is chosen which is most suitable for reconstructing the state (grey). Then, based on the already measured data, the active learner requests specific highly informative samples (blue). These are used to train a RBM representing the reconstructed quantum state.}
	\label{fig:AL}
\end{figure}

Here we show that it is possible to further reduce the number of required measurement samples by combining restricted Boltzmann machines with an adaptive active learning scheme as shown in Fig.~\ref{fig:AL}. Active learning (AL) machine learning models are able to interact with their environment. In the case of the QST scheme we propose here, this interaction comprises of requesting samples which are most informative for the further learning process \cite{Settles2009}. Active learning has been applied successfully, e.g. in improving classification tasks \cite{Greiner2002, TongChang2001} or speech recognition \cite{Tur2005}, in computational physics to speed up calculations of multidimensional functions \cite{Yao2020}, in condensed matter theory to map out a phase diagram \cite{Ding2022}, and in quantum experiment design \cite{Ding2020}. 
Here, we use AL in the sense that our model decides actively which measurement configuration to consider for consecutive measurements in QST to improve the training most efficiently. The set of measurements is hence adaptively optimized during the training, which has already been shown to bring advantages for full quantum state tomography of very small systems \cite{Huzar2012, Mahler2013, Ferrie2014, Straupe2016, Wu2020}. Existing protocols, however, are often not scalable to larger systems for various reasons, for example because either the reconstruction of the quantum state or the procedure to obtain the next measurement basis become numerically intractable. Another obstruction is the optimal measurement basis itself: some adaptive protocols are based on performing measurements along the direction of the current reconstruction, which can be highly entangled many-body states, and thus extremely challenging to implement. Our protocol overcomes these challenges by combining efficient state tomography based on neural networks with an active learning scheme relying on single qubit rotations only. We would like to point out that our method is designed to reduce the sampling complexity, but not the reconstruction complexity. The latter relies on classical resources, and is hence limited to quantum systems that can be simulated classically.  

In this article we show that active learning can strongly reduce the number of samples needed for the reconstruction compared to a passive procedure relying on RBMs only. We investigate quantum states consisting of up to $19$ qubits, or spin-$\frac{1}{2}$ particles, which are either generated on IBM quantum devices or simulated using the density-matrix renormalization group \cite{Schollwoeck2011}. Extensions of our AL scheme to other settings, such as fermionic systems or soft-core bosons, should be possible.

This article is organized as follows: we start with a brief summary of existing QST schemes by restricted Boltzmann machines, which constitute a building block of our improved active learning QST. We then introduce our active learning algorithm in section \ref{sec:Model}. In section \ref{sec:States}, we define and motivate the different quantum states considered in this work, in particular multi-qubit states with variable degrees of entanglement and ground states of a one-dimensional XXZ model and a kinetically constrained spin chain with a hidden U(1) symmetry. We present the results of our active learning scheme for quantum state reconstruction in section \ref{sec:Results} and conclude in section \ref{sec:Summary}. 

\section{Restricted Boltzmann Machines}

Since our AL scheme also includes the training of a committee of restricted Boltzmann machines, see Fig.~\ref{fig:AL}, we will introduce the latter with a focus on their application to QST in this section. We would like to emphasize that the idea of our AL scheme is independent of the specific representation of the reconstructed state and can in principle be applied in combination with any quantum state tomography scheme, such as other neural network architectures \cite{Rocchetto2018,Morawetz2021,Schmale2021,Ahmed2021,Cha2021} as well as matrix product state based state reconstruction \cite{Cramer2010,Baumgratz2013}. 

Here, we use the implementation of RBM quantum state tomography by Beach et al. \cite{Beach2019} in the form of the python package QuCumber. 
RBMs are capable of representing highly entangled quantum states \cite{Torlai2018,Gao2017} and can in principle be extended to deep Boltzmann machines to increase their expressivity \cite{Melko2019}. Furthermore, the RBM ansatz can be supplemented with an additional hidden layer to represent mixed states with arbitrary degree of mixedness, as shown in Refs. \cite{Torlai20182,Melko2019,Hendry2022}. In Ref. \cite{Torlai20182}, this \textit{purification scheme} has already been applied successfully to experimental measurements. Here, we restrict ourselves to the conventional RBM architecture representing pure states, with the limitations as discussed e.g. in Refs. \cite{Gao2017,Sehayek2019}.
Hereby, the target many-body quantum state is represented in terms of two RBMs with weight vectors $\lambda$ and $\mu$ which define the amplitude and the phase of the reconstructed state, i.e.
\begin{align}
\psi_{\mathrm{RBM}}(\bm{x})=\psi_{\lambda, \mu}(\bm{x}) = \sqrt{\frac{p_\lambda(\bm{x})}{Z_\lambda}}\mathrm{e}^{i\theta_\mu(\bm{x})/2}\, ,
\label{eq:RBMstate}
\end{align}
where $\bm{x}$ labels a general set of basis states. Details on the RBM wave function can be found in Appendix \ref{appendix:RBM}. 
\begin{figure}[t!]
	\centering
  \includegraphics[width=0.45\textwidth]{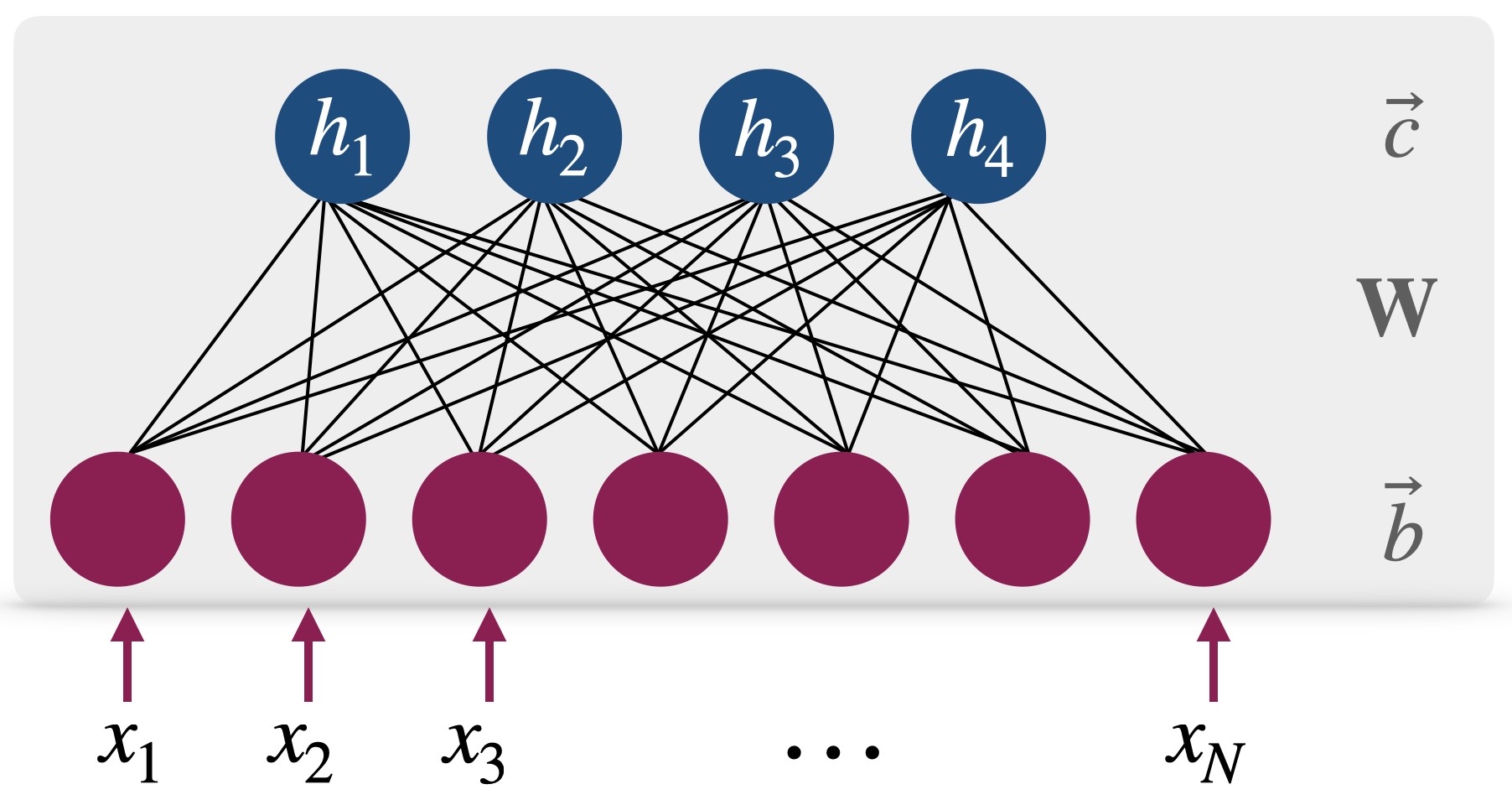}
	\caption{Schematic representation of an RBM with visible nodes $\vec{x}$, hidden nodes $\vec{h}$, and weights $\vec{b}$, $\vec{c}$ and $\bf{W}$.}
	\label{fig:RBM}
\end{figure}

A RBM consist of layers of so-called visible and hidden nodes $v_i$ and $h_j$ with bias weights $b_i$ and $c_j$. A schematic representation is shown in Fig. \ref{fig:RBM}. The visible nodes correspond to the input data, consisting of measurements of the state under investigation and given in terms of bit strings of zeros and ones. Both layers are connected and each connection is weighted by parameters $W_{ij}$. To reconstruct a state and represent it in terms of a RBM, the network parameters are learned from a set of measurements. To this end, the weights $\bm{b}, \,\bm{c}$ and $\bm{W}$ are adjusted such that the Kullback-Leibler (KL) divergence,
\begin{align}
KL(q, p_{(\lambda, \mu)}) = \sum_{\bm{x} \in \mathcal{D}}q(\bm{x})\mathrm{log}\frac{q(\bm{x})}{p_{(\lambda, \mu)}(\bm{x})},
\label{eq:KL}
\end{align}
is minimized\footnote{The empirical distribution $ q(\bm{x})$ is defined as follows: if a possible outcome $\bm{v}$ is contained $N_{\bm{v}}$ times in the set of $N_{\rm tot}$ existing measurements $\mathcal{D}$, then $q(\bm{v}) = N_{\bm{v}}/N_{\rm tot}$. If $\bm{x} \notin \mathcal{D}$ has not been measured, $q(\bm{x})=0$.}, which quantifies how close the reconstructed distribution $p_{(\lambda, \mu)}=\vert \psi_{\mathrm{RBM}}(\bm{x})\vert ^2$ is to the measured distribution $q(\bm{x})$. Details on the training procedure can be found in Appendix \ref{appendix:RBM}.

While the Kullback-Leibler divergence can be determined without prior knowledge of the target state\footnote{In our convention, the target state is the state that would be prepared on an ideal quantum device. For real quantum devices as used later, this may differ from the state that was actually prepared.}, a commonly used way of benchmarking of a QST scheme is a direct comparison with the target state. In order to do so, a useful measure for the performance of the reconstruction is the fidelity
\begin{align}
    f = \vert \langle \psi_\mathrm{RBM} \vert \psi_\mathrm{target} \rangle \vert ^2,
    \label{eq:Fidelity}
\end{align}
which represents the square of the overlap of the RBM state representation and the target state \cite{QucumberDoku}. Note that the fidelity can only be evaluated if the target state vector is explicitly known. For multi-qubit systems, $f$ is often re-scaled to $\tilde{f} = f^\frac{1}{N}$, with $N$ the number of qubits, to account for the exponential size of the underlying Hilbert space. Another way to benchmark the performance of the reconstruction scheme is to compare observables like the density or correlators of the target and reconstructed states, which will be done in section \ref{sec:DMRG}.\\

To obtain information about both the amplitude and phase of the target state, different types of measurements are required: 
\begin{itemize}
    \item(i)] Samples from a \textit{reference} measurement configuration $C_\mathrm{ref}$ are needed to learn the wave function amplitudes from the measurement outcomes.
    \item[(ii)] Samples from measurements in further, different basis configurations are required to extract information about the phase. Each measurement in a different basis configuration corresponds to a rotation of one or several qubits into the $x$, $y$ or $z$ basis individually (see Appendix ~\ref{appendix:RBM} for the definition of the rotation matrices). 
\end{itemize} 
In the following, we will denote measurement configurations by $C$ and the rotation to the respective measurement configuration by $R_C$.

\section{Active Learning Algorithm \label{sec:Model}}

Depending on the properties of the quantum state under consideration, there are different configurations which contain relevant information and are hence most useful for the reconstruction \cite{Mahler2013}. For most states it is difficult and time-consuming to obtain the optimal set of measurement configurations and amount of samples which leads to a good reconstruction (see Appendix \ref{appendix:IBM}). Here, we apply AL to choose the configurations during the learning process that contain most new pieces of information about the state under consideration. Note that although the choice of quantum state representation in terms of an RBM comes with the limitations discussed above, our AL scheme itself does not make any assumptions on the type of state or its properties and is not restricted to the specific choice of neural wave function representation. Within our scheme, the configurations are chosen such that the amount of information contained in a finite number of measurements from this configuration is larger than for other configurations. We show that by requesting this specific, highly informative data, the total number of samples can be reduced while the accuracy of the machine learning model is increased compared to a procedure which uses RBMs only. 

\begin{figure}[t]
	\centering
  \includegraphics[width=0.495\textwidth]{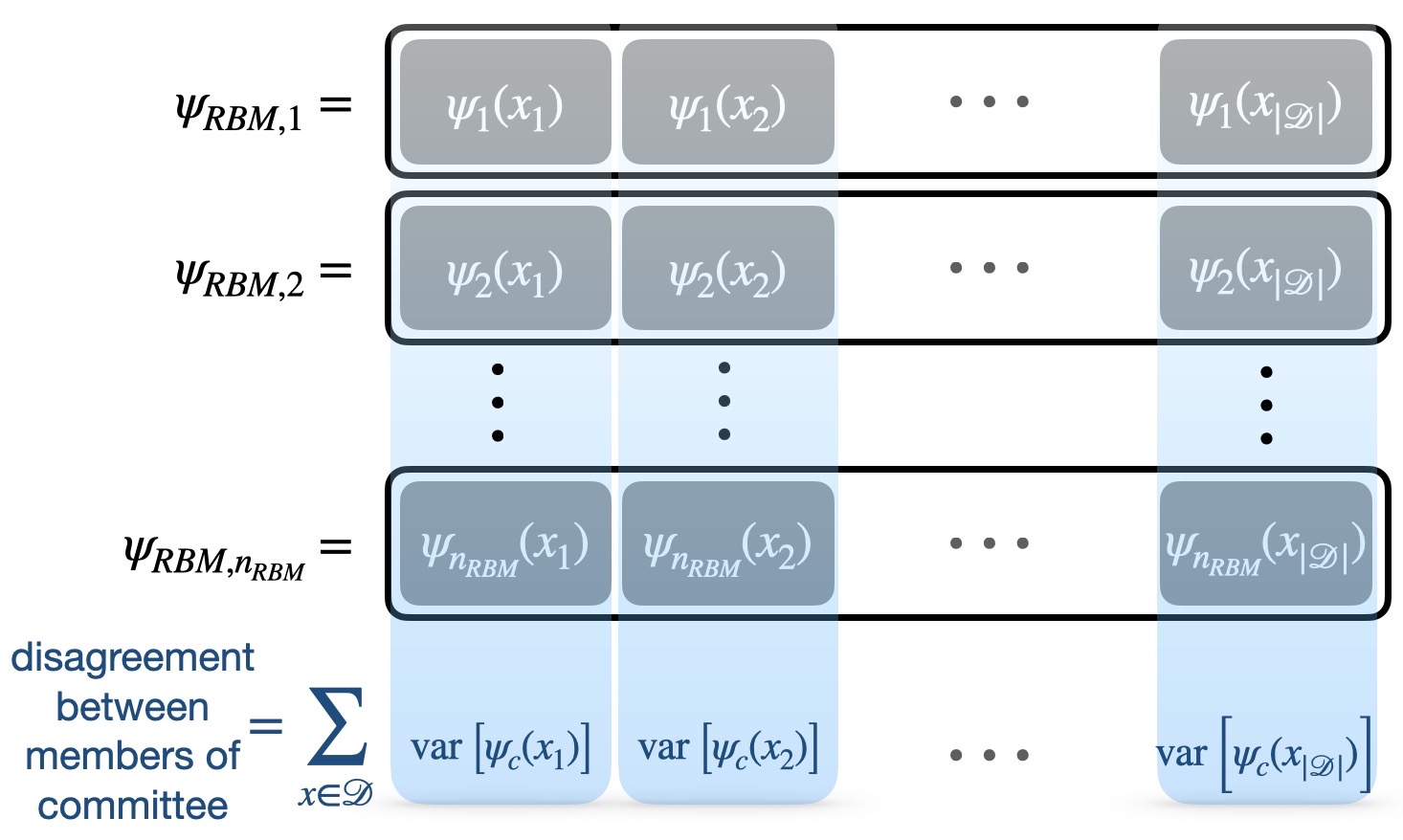}
	\caption{Calculating the level of disagreement between different wave function representations of different members of the committee (rows) as given by Eq. \eqref{eq:disagreement}. The shown calculation is performed separately for each configuration $x_i$. For simplicity, we show the procedure for real-valued wave functions and refer the reader to the text for the explanation of the full AL scheme that includes the states' phase structure.}
	\label{fig:disagreement}
\end{figure}

The choice for the next measurement configuration is the core of our active learning algorithm. It is made by applying the so-called \textit{query-by-committee} strategy from active learning theory \cite{Freund1997,Seung1992,Settles2009}. Hereby a committee of $n_\mathrm{RBM}$ models (here RBMs) $\Theta=\{\theta_1, \dots \theta_{n_\mathrm{RBM}}\}$, initialized with different parameters, is trained on the same data. In each active learning step, each member of the committee $\Theta$ casts its vote on how the true wave function should look like, based on their learned quantum state representations $\{\psi_{\mathrm{RBM},1},\dots,\psi_{\mathrm{RBM},n_\mathrm{RBM}}\}$. 
The committee selects the measurement configuration $C^{*}$ for which all models disagree most on the training outcomes $\{R_C\psi_{\mathrm{RBM},1},\dots,R_C\psi_{\mathrm{RBM},n_\mathrm{RBM}}\}$. As shown in Ref. \cite{Seung1992}, the degree of disagreement among the committee members can be seen as an estimate for the information value.\\
A common strategy to calculate the level of disagreement is the \textit{Kullback-Leibler divergence to the mean} \cite{McCallum1998,pereira1993}, which measures the distance between the probability distributions $\{P_{\theta_1},\cdots ,P_{\theta_{n_{RBM}}}\}$ based on the Kullback-Leibler \eqref{eq:KL} divergence as a distance measure between the probability distributions of each model, $P_{\theta}$, and the average distribution of all models $P_\Theta$ with $P_\Theta(\mathbf{x})=\frac{1}{n_\mathrm{RBM}}\sum_{\theta \in \Theta} P_{\theta}(\mathbf{x})$. Then, the configuration $C^{*}$ that maximizes the distance is selected. Here, we adopt the same strategy, but with a different distance measure given by the \textit{Jeffrey's} distance \cite{Chung1989}
\begin{align}
    J(P_{\theta},P_\Theta)=\sum_{\mathbf{x}\in \mathcal{D}}\left(\sqrt{P_{\theta}(\mathbf{x})}-\sqrt{P_\Theta(\mathbf{x})}\right)^2
\end{align} 
instead of the $KL$ to measure the distance between the probability distributions $\{P_{\theta_1},\cdots ,P_{\theta_{n_{RBM}}}\}$. In contrast to the $KL$ the Jeffrey's distance has symmetric contributions for both $P_\theta(\mathbf{x})\leq P_\Theta(\mathbf{x})$ and $P_\theta(\mathbf{x})\geq P_\Theta(\mathbf{x})$ since no logarithm is involved, ensuring that all members of the committee have equal votes. More precisely,
\begin{align}
    C^{*}=\argmax_C \, \frac{1}{n_\mathrm{RBM}}\sum_{\theta\in\Theta}J(P_{\theta},P_\Theta)\,
\end{align}
which can be brought to the form
\begin{align}
    \frac{1}{n_\mathrm{RBM}} &\sum_{\theta\in\Theta}J(P_{\theta},P_\Theta)\notag\\
    &= \frac{1}{n_\mathrm{RBM}}\sum_{\theta\in\Theta}\sum_{\mathbf{x}\in \mathcal{D}}(\sqrt{P_{\theta}(\mathbf{x})}-\sqrt{P_\Theta(\mathbf{x})})^2\notag \\
    &=\sum_{\mathbf{x}\in \mathcal{D}} \mathrm{var}\left[\sqrt{P_{\{\theta\}}(\mathbf{x})}\right] \, .
    \label{eq:disagreement}
\end{align}
with the empirical variance $\mathrm{var}\left[\sqrt{P_{\{\theta\}}(\mathbf{x})}\right] = \frac{1}{n_\mathrm{RBM}}\sum_{\theta \in \Theta} (\sqrt{P_{\theta}(\mathbf{x})}-\sqrt{P_\Theta(\mathbf{x})})^2 $ of the wave functions amplitude $\vert \psi_\mathrm{RBM} (\mathbf{x})\vert = \sqrt{P_{\theta}(\mathbf{x})}$. The calculation of this term is shown in Fig. \ref{fig:disagreement}. It has been shown that query-by-committee algorithms work in some cases already for very small committees with only up to three members \cite{Seung1992,McCallum1998,Settles2009}. In order to take complex wave functions into account, we calculate Eq. \eqref{eq:disagreement} for the phase part of $\psi_\mathrm{RBM}$ as well and normalize the amplitude and phase variance by the respective $L1$ normalized values. Depending on the size of the Hilbert space, the calculation can be done by comparing the full state vectors (i.e. calculating the variance of $\psi_\mathrm{RBMi}(\bf{x})$ for \textit{every} basis state $\bf{x}$) or by sampling from each of the $n_{\mathrm{RBM}}$ RBM distributions. \\

A schematic overview of our complete active learning algorithm is provided in Fig.~\ref{fig:AL}. It consists of the following steps: 

\begin{itemize}
    \item[1.)] First, samples in only three different measurement configurations are generated from the experiment: without rotations (denoted by $C_1=(zz\dots z)$), with all qubits rotated to the $x$ axis ($C_2=(xx\dots x)$) and, third, with all qubits rotated to the $y$ axis ($C_3=(yy\dots y)$). We denote the respective measurement sets by $\{\bm{x}_{z\dots z}\}$, $\{\bm{x}_{x\dots x}\}$ and $\{\bm{x}_{y\dots y}\}$. $C_i$ are the candidates for the reference configuration $C_{\mathrm{ref}}$ that is selected in the next step.
    \item[2.)] The reference basis $C_{\mathrm{ref}}$ for the reconstruction of the quantum state's amplitude is selected among the candidates from step 1, based on the training outcome of the RBM members of the committee on measurements $\{\bm{x}_{z\dots z}\}$, $\{\bm{x}_{y\dots y}\}$ and $\{\bm{x}_{y\dots y}\}$. The details of this step are described at the end of this section. The measurements corresponding to the selected reference configuration will be used in the next steps. 
    \item[3.)] Third, we
    train $n_{\mathrm{RBM}}$ different RBMs with different weight initializations on the data in the pool of samples by minimizing the $KL$ divergence \eqref{eq:KL}. Note that this step can easily be parallelized.
    \item[4.)] The active learner requests new measurements from a specific measurement configuration $C^{*}$. Here, $C^{*}$ can consist of any combination of local rotations to the $x$, $y$ or $z$ basis and hence $3^N$ rotations (denoted by $\{R_{C}\}$) come into question. We apply the \textit{query-by-committee} strategy as described above to allow an efficient choice of $C^{*}$. To this end, we rotate the RBM representations using the rotations $\{R_{C}\}$ and request samples from those configurations for which the RBMs disagree most. The level of disagreement is calculated using eq. \eqref{eq:disagreement} for the amplitudes and phases separately. In order to make them comparable, we normalize by the absolute ($L1$ normalized) value of the amplitudes and variances, respectively. Note that in this step the rotations $\{R_{C}\}$ will be applied to the RBM representations of the wave functions and not to the actual quantum state, which can be done in a much more efficient way.
    \begin{itemize}
        \item [a)]If the RBMs disagree more on the amplitudes than on the phases of the reconstructed states (i.e. the variance of the amplitudes is larger than the variance of the phases of $\psi_{\mathrm{RBM}}$ for different RBMs), we skip the full \textit{query-by-committee} procedure and request measurements from the reference basis.
        \item[b)] Else, the probability distributions $p_{(\lambda_i, \mu_i)}(\bm{x})=\vert \psi_{\mathrm{RBM}i}(\bm{x})\vert ^2$ with $\psi_{\mathrm{RBM}i}$ rotated to $c\leq 3^N$ different configurations for all $i = 1, \dots, n_{\mathrm{RBM}}$ RBMs are used to determine the best choice for the next configuration $C^{*}$ as described above (with $N$ the system size). 
    \end{itemize}
    \item[5.)] Samples in the requested configuration are measured and steps 3 and 4 are repeated.
    \item[6.)] The training is stopped when the error of observables from the target and RBM wave functions drop below a pre-defined threshold value or when a maximum number of queries is reached. The stopping criteria for the states under consideration are specified in the respective sections. 
\end{itemize}
Furthermore, our AL-QST scheme automatically requests samples from the two other configurations beside the reference configuration mentioned in 1 if the reference configuration is requested two times in a row. For example, if the reference configuration is the $zz\dots z$ configuration and measurements from this configuration are requested two times in a row, samples from the $xx\dots x$ and $yy\dots y$ configurations will be added in the next learning cycles. This is particularly relevant for rotationally invariant systems like the Heisenberg chain. Hereby, no new measurements are needed since the discarded measurements from 1 can be used. \\

Note that step 4b is a bottleneck of our approach, since the computational effort to calculate the variance for all possible configurations scales exponentially with $N$. However, this scaling can still be favorable for relative large system sizes since it only concerns the RBM representation of the target wave function, and not the number of measurements or measurement configurations from the experiment. Furthermore, we show that our AL scheme still improves the reconstruction when it is applied to a randomly chosen subset of all possible configurations. Potential ways to overcome the exponential scaling are discussed at the end of this paper. \\

The reference measurement configuration chosen in step 2 is selected in a similar way as the new configurations are chosen in step 4. However, the important difference to the \textit{query-by-committee} strategy as in step 4 lies in the fact that a good reference configuration $C_{\mathrm{ref}}$ should be the best available staring point for the training procedure. Hence, we want to select $C_i$ with the best estimate of the target wave function. In a similar spirit as in step 4, we train all members of the committee, but on each data set $\{\bm{x}_{z\dots z}\}$, $\{\bm{x}_{x\dots x}\}$ and $\{\bm{x}_{y\dots y}\}$ obtained in step 1 separately and, in contrast to step 4, we choose the candidate $C_{i}$ where the $n_{\mathrm{RBM}}$ RBMs \textit{agree most} on the wave function state vectors, i.e. where the variance is the lowest. \\

The reference to an implementation of our algorithm is provided at the end of this paper.

\section{Considered States \label{sec:States}}
Before we present results by our AL-QST scheme, we provide an overview of the quantum states used to benchmark our method. We chose two types of states: first a set of generic qubit states with variable degree of entanglement, which provide direct insights into the performance of our scheme. Second, we consider many-body ground states of one-dimensional spin Hamiltonians as an illustration of our method for quantum simulators. 

\subsection{Qubit states and IBM's Quantum cloud}

We investigate the reconstruction of states which are generated on real quantum devices and classical simulators of the \textit{IBM Quantum} platform \cite{IBM}. IBM provides 21 quantum systems based on superconducting qubits which can be accessed via a cloud. Systems with up to five qubits are accessible for non-internal users with a free account. Using Qiskit it is possible to implement quantum algorithms on these quantum systems \cite{Qiskit2010}.

Here, we consider the reconstruction of four target states:
{Greenberger–Horne–Zeilinger} (GHZ) states, 
    \begin{align}
    \ket{\mathrm{GHZ}} = \frac{1}{\sqrt{2}} \left(\ket{0\dots 0} + \ket{1\dots 1} \right),
    \label{eq:State1}
    \end{align} 
GHZ states with a \textit{complex phase}, 
    \begin{align}
    \ket{\mathrm{GHZ}_{\varphi}} = \frac{1}{\sqrt{2}} \left(\ket{0\dots 0} + i\ket{1\dots 1} \right),
    \label{eq:State1_phase}
    \end{align} 
polarized product states with all qubits set to one (spin chains with spins pointing in $z$ direction),
\begin{align}
    \ket{z\mathrm{-spins}} = \ket{1\dots 1},
    \label{eq:State2}
\end{align}
and states with a state vector with equal amplitudes for all components. They correspond to spin chains with all spins pointing in $x$ direction, e.g. for two qubits
\begin{align}
    \ket{x\mathrm{-spins}} = \frac{1}{2} \ket{00}+  \frac{1}{2} \ket{01} +  \frac{1}{2}   \ket{10} +  \frac{1}{2}  \ket{11}.
    \label{eq:State3}
\end{align}

Using the IBM platform we investigate the reconstruction of states with five qubits. Tools for rotating and measuring the quantum systems are provided by Qiskit. An example for measuring a two-qubit system in the $xy$ basis (first qubit rotated to the $x$ axis, second qubit rotated to $y$) is shown in Appendix \ref{appendix:RBM}. In the following sections we will use the same notation, where e.g. $xx\dots x$ denotes a system with all qubits rotated from the $z$ to the $x$ axis.


\subsection{XXZ and Heisenberg states}
As a second example, we use our AL-QST scheme to reconstruct ground states of the XXZ Hamiltonian
\begin{align}
\hat{\mathcal{H}} = \sum_{\langle i,j\rangle} J (\hat{S}_i^x \hat{S}_j^x+\hat{S}_i^y\hat{S}_j^y)+J(1+\Delta)\hat{S}_i^z\hat{S}_j^z\,,
\label{eq:Heisenberg}
\end{align}
where $\hat{S}_{i(j)}^{\mu}$ denotes the $\mu\in \{x,y,z\}$ component of the spin-1/2 operator at site $i(j) =1,\dots ,L $. This model encloses ground states with a strong polarization in $z$ direction for large 
$\vert \Delta\vert$ (broken \Zt symmetry) to critical states without 
long-range order for $-2J \leq \Delta\leq 0$. For $\Delta=0$ the ground states are SU(2) invariant (Heisenberg antiferromagnet). In what follows we will compute the ground states of Eq.~\eqref{eq:Heisenberg} using a matrix product state representation with the SyTen package \cite{syten1, syten2}.

\subsection{Kinetically constrained spin chain}
As an additional illustrative example, we apply AL-QST to reconstruct ground states of a kinetically constrained one-dimensional spin chain (KCS) model. It is described by the following Hamiltonian \cite{Iadecola2020,Borla2020,Kebric2021}:
\begin{align}
    \H = t \sum_{j=2}^{L-1} &\l 4 \hat{S}^{x}_{j-1} \hat{S}^{x}_{j+1} - 1 \r \hat{S}^{z}_j \nonumber \\
    &- h \sum_{j=1}^{L} 2 \hat{S}^{x}_{j}
    + \mu \sum_{j=2}^{L} \hat{S}^{x}_{j-1} \hat{S}^{x}_{j},
    \label{eq:LGT_Model_Spin_Def}
\end{align}
where $\hat{S}^\mu_j$ denotes a spin-$1/2$ operator ($\mu=x,y,z$) on site $j=1,...,L$. This model has a very interesting interpretation, where the spin domain walls in the $x$-basis correspond to particles on a dual lattice.
More precisely, model \eqref{eq:LGT_Model_Spin_Def} can be mapped to a one-dimensional \Zt lattice gauge theory with matter \cite{Borla2020}, see Appendix \ref{ApdxKinConsSpn}.

We choose the system in Eq.~\eqref{eq:LGT_Model_Spin_Def} due to several non-trivial properties for which we can check in the reconstructed state. These include: (i) that the underlying excitations are domain walls of the spins, extending beyond one site -- a fact that needs to be captured by a reliable QST scheme; (ii) the Hamiltonian features a hidden U(1) symmetry describing the conservation of the total number of domain walls -- their number is controlled by the chemical potential $\mu$; (iii) the model hosts gapless Luttinger liquids with significant amount of non-local entanglement, presenting a general challenge for any QST scheme; and (iv) by tuning the effective field from $h=0$ to $h \neq 0$, a confinement-deconfinement transition exists where the nature of constituents of the Luttinger liquid changes, as indicated by a change of the Fermi momentum $k_{\rm F}$ and reflected in the period of Friedel oscillations \cite{Borla2020,Kebric2021}.

Below we compute the ground states of Eq.~\eqref{eq:LGT_Model_Spin_Def} with the density-matrix renormalization group (DMRG) using the SyTen package \cite{syten1,syten2}. We obtain a MPS representation of the ground states, from which efficient snapshot sampling is possible \cite{Ferris2012}, including in variable bases \cite{buser2021arXiv}. This provides us with the data required to run and benchmark the AL-QST algorithm.

To probe how close the reconstructed state is to the actual ground state, we compute the following observables. We start with the local domain-wall density,
\begin{equation}
    \hat{n}(j) = \frac{1}{2}
    \l 1 - 4 \hat{S}^{x}_{j}\hat{S}^{x}_{j+1} \r.
    \label{eq:LocalDensitySpinDef}
\end{equation}
This also immediately leads us to the conserved total system density $\hat{n}^{\rm tot} = \frac{1}{L-1} \sum_{j=1}^{L-1} \hat{n}(j)$. In practice, we find it convenient to define a vector $\bm{n}$ of local densities with the following entries,
\begin{equation}
    \bm{n} = \bigl( \langle \hat{n}_{1} \rangle, \langle \hat{n}_{2} \rangle, ..., \langle \hat{n}_{L} \rangle \bigr)^T.
\end{equation}

In order to probe the confinement of domain walls we consider their non-local equal-time Green's function measured relative to the center $L/2$ (for $L$ odd: $L/2+1/2$) of the chain (see Appendix \ref{ApdxKinConsSpn} for more details):
\begin{multline}
    c(d) = \left \langle
    \frac{1}{2} \l 1 - 4\hat{S}^{x}_{L/2}\hat{S}^{x}_{L/2+1} \r \l \prod_{L/2 < j \leq L/2+d}
    2 \hat{S}^{z}_{j} \r  \right. \\
   \times \left. \frac{1}{2} \l 1- 4\hat{S}^{x}_{L/2+d}\hat{S}^{x}_{L/2+1+d} \r
    \right \rangle .
    \label{eq:GreensSpinDef}
\end{multline}
We highlight the following key properties of this function: (a) for distance $d=0$, the density is recovered, $c(0) = \langle \hat{n}(L/2) \rangle$; (b) the decay of $c(d)$ allows to distinguish between confined (exponential decay) and deconfined (power-law decay) regimes.

\section{Results\label{sec:Results}}

\begin{figure}[t]
	\centering
  \includegraphics[width=0.48\textwidth]{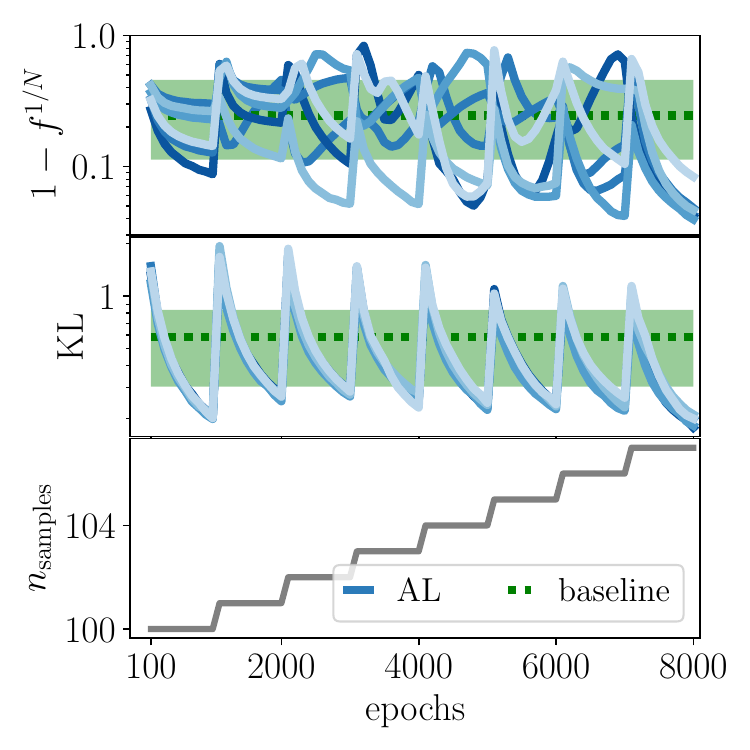}
	\caption[Learning curve with active learning]{Learning curve: re-scaled fidelities (top), Kullback-Leibler divergence (middle) and number of samples within the learning cycles (bottom) with active learning for a GHZ state with phase structure and consisting of $5$ qubits. Here, $6$ different RBMs are used in step 4 of the active learning algorithm (blue lines). At the end, $107$ samples were used. The results can be compared to a learning cycle without active learning (green). The green band corresponds to the entire span of results without AL. The active learner decided to choose the $zz\dots z$ basis as reference basis in step 2 and requested measurements from $2$ additional configurations in $7$ queries (see epoch s $1000$ to $8000$ and step 4 of the AL-QST learning scheme).}
	\label{fig:ALLearningCurve}
\end{figure}

The active learning curve for an exemplary state, namely a GHZ state with phase structure consisting of $5$ qubits and generated with the IBM quantum simulator, is shown in Fig.~\ref{fig:ALLearningCurve}. The results for AL tomography are compared to QST without active learning (denoted \textit{baseline}) by using the pure RBM reconstruction of QuCumber with equally many, but randomly chosen measurement configurations, and the same number of samples as for AL (without taking the discarded data from step 1 into account in both cases). 

Note that we consider the number of samples and configurations at the end of the AL scheme for the baseline reconstruction. Hence, the results for AL and baseline can differ already at the beginning of the training. The pool of samples which is used for the RBM training is the same for all RBMs (both for AL and baseline, respectively).\\

Fig.~\ref{fig:ALLearningCurve} shows the learning curve for the AL scheme as presented in Fig.~\ref{fig:AL}. After the selection of the reference configuration ($zzzzz$, not shown in Fig.~\ref{fig:ALLearningCurve}) the training starts with $100$ samples drawn from the reference basis, which are fed into the QuCumber state reconstruction as implemented by Beach et al. \cite{Beach2019} (see step 3). In the subsequent learning sequence up to epoch $1000$, the re-scaled fidelity $f^{1/N}$ increases ($1-f^{1/N}$ decreases) and the Kullback-Leibler divergence $KL$ decreases. However, the samples measured in step 1 of the active learning routine do not contain enough information to decrease $1-f^{1/N}$ below a threshold value of $10\, \%$ for all RBMs (see first section in Fig.~\ref{fig:ALLearningCurve}). The lack of information is mostly due to the fact that no information on the phase structure of the state under consideration is contained in the measurements from the $zzzzz$ configuration. 

After that, more and more samples are added to the pool of measurements which are requested by the learner and the RBM learning process is re-started (with the same random initialization of the RBMs as before). For the state reconstruction presented in Fig.~\ref{fig:AL}, the learner requests a measurement from the $xzzxx$ configuration in the first $6$ queries (see epochs $1000$ to $6000$), and from $zyyyz$ in the last learning cycle starting in epoch $7000$. Note that samples from these measurement configurations contain not only information on the amplitude, but also on the phase structure of the state under investigation. Step by step, these samples are added to the learning process. This process is stopped when a specified re-scaled fidelity (here $f^{1/N}_{\mathrm{stop}}=90\, \%$) is reached or the number of posed queries exceeds a maximal value $N_{\mathrm{query}}^{\mathrm{max}}$ (in this paper we use $N_{\mathrm{query}}^{\mathrm{max}}=30$). The final fidelity in this example is $f^{1/N}=95.0 \pm 1.5
\,\%$ (average over the six RBMs used for step 3) using only $107$ samples and three different configurations. Note that we restart the learning from randomly initialized weights in each AL step and not from the trained models from the first steps to avoid a bias from the incomplete data, yielding the periodic peaks in fidelity and KL.
Furthermore, it can be observed that there are regimes in the learning procedure where the difference between reconstructed and target state stagnates or even grows when adding the requested samples. This is not unintentional within our learning scheme: If a new sample is requested that contains completely different information (i.e. when adding the first sample from the $xzzxx$ configuration to a pool of $100$ samples from the $zz\dots z$ configuration in epoch $1000$) the new information cannot be successfully incorporated into the context of the data from previous AL steps in the first place. However, when requesting more and more samples afterwards, the same piece of information can become a valuable contribution to a good reconstruction.

A learning procedure without AL with the same number of measurements and equally many, but randomly chosen configurations (which will be called \textit{baseline} in the following) ends with $f^{1/N}_{\mathrm{baseline}}=75\pm 11\,\%$ (average over all RBMs, trained with the same samples), indicated by the green dotted line in Fig.~\ref{fig:ALLearningCurve}. Hence, the quality of the reconstruction was improved strongly by the use of active learning. Furthermore, the span of final fidelities for different RBMs is significantly decreased compared to a learning procedure without active learning (visualized by the green band in Fig.~\ref{fig:ALLearningCurve}), which makes the results much more reliable than without AL. Note moreover that besides $1-f^{1/N}$ also the $KL$ is reduced compared to the baseline, i.e. the final RBM reconstructions using AL can capture the given dataset in a more efficient way than without AL.\\

In the following sections, the results for the active learning procedure as explained in section \ref{sec:Model} will be presented for states on IBM's quantum devices and states generated with DMRG. The number of samples $N_{\mathrm{tot}}$, number of queries $N_{\mathrm{queries}}$, number of samples per query $N_{\mathrm{per\,query}}$ and number of different configurations $N_{\mathrm{conf}}$ are presented in Tab.~\ref{tab:Samples_IBM} and \ref{tab:Samples_DMRG}. Based on our experience that for a good reconstruction of the amplitude in most cases more information about the reference configuration is needed (compared to the reconstruction of the phase based on locally rotated configurations), we request more samples than $N_{\mathrm{per\,query}}$ (if not stated differently $3\cdot N_{\mathrm{per\,query}}$) if the reference configuration is requested.

\subsection{Tomography results for quantum states on IBM's Quantum cloud \label{sec:States_IBM}}

\begin{figure}[t]
	\centering
  \includegraphics[width=0.48\textwidth]{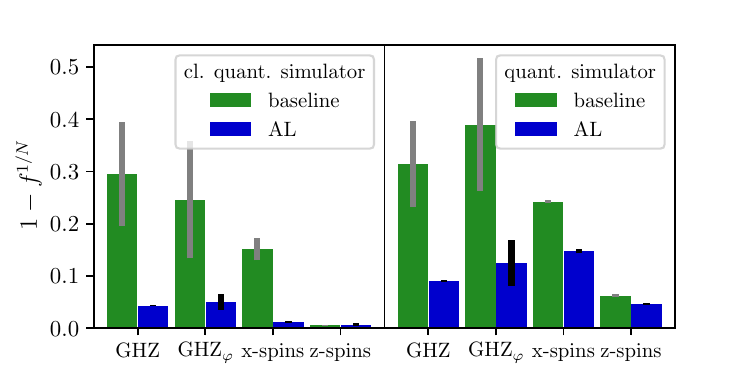}
	\caption[]{Final re-scaled fidelities for the states defined in Eqs.~\eqref{eq:State1}, \eqref{eq:State1_phase}, \eqref{eq:State2} and \eqref{eq:State3} with $5$ qubits generated on classical quantum simulators (left) and real quantum devices (right). An exemplary learning curve can be found in Fig.~\ref{fig:Example_GHZIBMclassical}.}
	\label{fig:IBM}
\end{figure}

In this section the results for states defined in Eqs. \eqref{eq:State1} to \eqref{eq:State3} generated on IBM's classical quantum simulator, which is designed to  mimic the execution of an actual device, and on real quantum devices are presented (for more details see appendix~\ref{appendix:IBM}). We consider a system size of $5$ qubits. The reconstruction results can be found in Fig.~\ref{fig:IBM}. Here the training is stopped as soon as $f^{1/N}_{\mathrm{stop}}=90\,\%$ is reached for the classical quantum simulators. For the real devices, we calculate the fidelity w.r.t. a perfect target state, although it may differ from the actually prepared state on the device. Consequently, the reconstruction fidelity is limited by the preparation errors and we set $f^{1/N}_{\mathrm{stop}}$ to a lower value $f^{1/N}_{\mathrm{stop}}=80\,\%$ (except for the GHZ state with $f^{1/N}_{\mathrm{stop}}=90\,\%$ and the $\mathrm{GHZ}_\varphi$ state with $f^{1/N}_{\mathrm{stop}}=85\,\%$). The number of samples, queries and configurations at the end of the training is presented in Tab.~\ref{tab:Samples_IBM}.


When reconstructing the GHZ state with AL, the active learner selects the $zzzzz$ basis as the reference basis. The fidelity of the GHZ classical quantum simulator states is improved by around $25\,\%$ from $f^{1/N}=70.49\pm0.10\,\%$ without AL to $f^{1/N}=95.667\pm0.001\,\%$. For real quantum devices the quality of the reconstruction improves as well, by around $22\,\%$ from $f^{1/N}=68.6\pm8.3\,\%$ without AL to $f^{1/N}=90.93\pm0.16\,\%$ with respect to the theoretically ideal state. Moreover, the standard deviation of the results for different RBMs is lowered by up to two order of magnitude for the classical quantum simulators and quantum devices, which makes the results more robust when using AL. This improvement is achieved by requesting samples from only two additional measurement configurations.

\begin{figure}[t]
	\centering
  \includegraphics[width=0.48\textwidth]{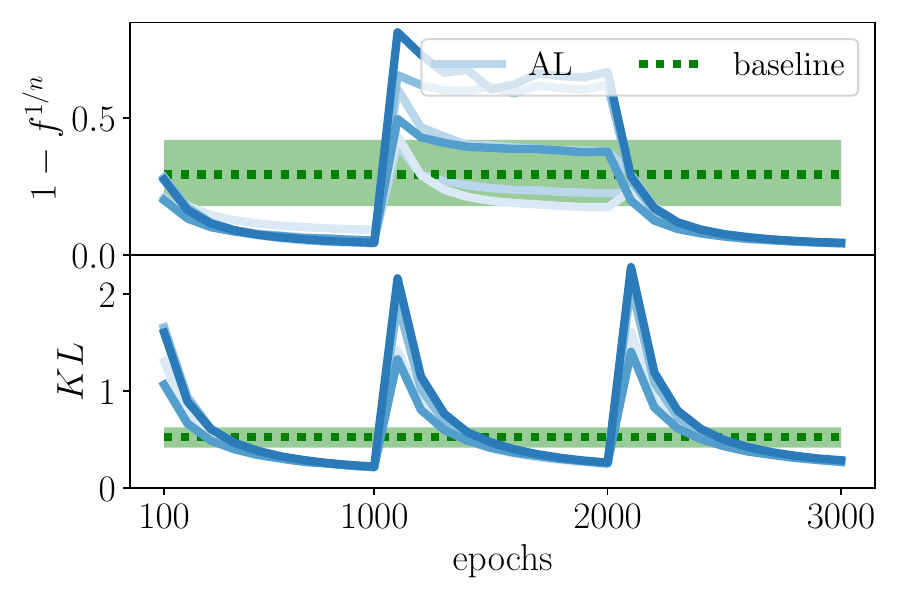}
	\caption[]{Exemplary learning curves for the GHZ state (see Eq.~\eqref{eq:State1}) with $5$ qubits generated on a classical quantum simulator by IBM. Both $1-f^{1/N}$ (with $f^{1/N}$ the fidelity) and the Kullback-Leibler divergence $KL$ are reduced when using AL, compared to a reconstruction with the same number of measurements, but randomly chosen measurement configurations.}
	\label{fig:Example_GHZIBMclassical}
\end{figure}

In Fig.~\ref{fig:Example_GHZIBMclassical} the learning curve for the GHZ state is shown. When using AL the fidelity and $KL$ divergence are decreased strongly at the end of the training. One can see that in the first learning cycle, which uses $60$ samples drawn from the $zzzzz$ reference basis, the information contained in these samples is not enough to decrease the fidelity below the threshold value of $f^{1/N}_{\mathrm{stop}}=90\,\%$ for most RBMs. By adding only one sample of the $yzzzy$ configuration to the pool of samples in the first query, the RBMs cannot successfully put the piece of new information in the context of the old measurements at first and and the infidelity increases compared to the first learning step. However, one can see that the RBMs still adapt to the new information since the $KL$ is decreased to a comparable amount as in the first part of the learning. In the second query one sample from $xxyxy$ is added. Together with the other samples, the information contained in the measurements is enough to adapt the phase of the reconstructed wave function such that  the fidelity increases above the threshold value. \\

\begin{figure}[t]
	\centering
  \includegraphics[width=0.48\textwidth]{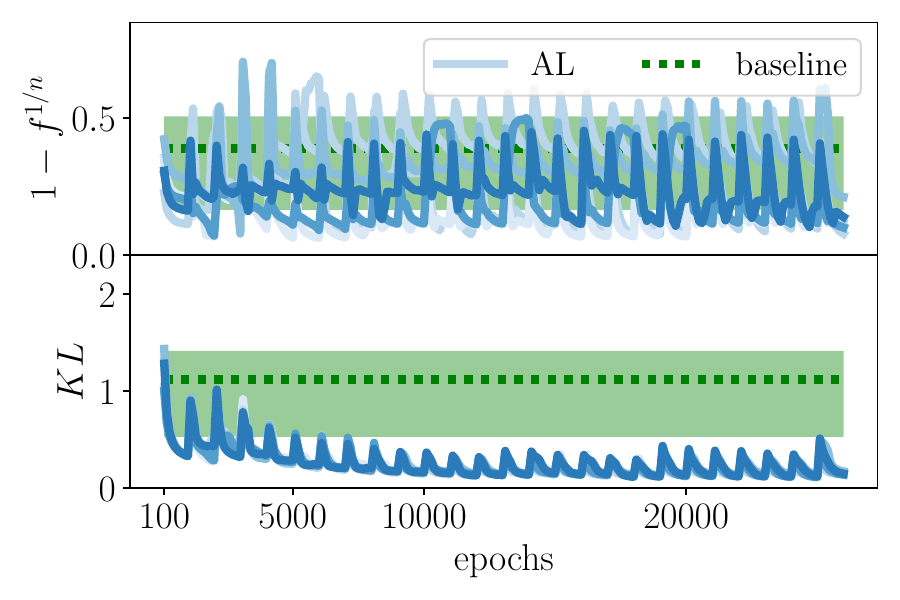}
	\caption[]{Exemplary learning curves for the GHZ state with phase (see Eq.~\eqref{eq:State1_phase}) with $5$ qubits generated on a quantum device of IBM. Both $1-f^{1/N}$ (with $f^{1/N}$ the fidelity) and the Kullback-Leibler divergence $KL$ are reduced when using AL, compared to a reconstruction with the same number of measurements, but randomly chosen measurement configurations.}
	\label{fig:Example_GHZphase_IBMquantum}
\end{figure}

For the GHZ state with phase structure the AL-QST scheme also requests additional measurements in 2 (5) additional measurement configurations for states generated with the classical simulator (real quantum device). In contrast to the GHZ state without phase structure, much more information is contained in these configurations, namely the phase difference of $\ket{0\dots0}$ and $\ket{1\dots1}$. Our scheme is able to capture the importance of measurements from other configurations than the reference configuration and requests many of the configurations several times until the the learning is stopped after 7 (25) queries. The AL-QST scheme improves the reconstruction results to a fidelity of $f^{1/N}=95.0\pm 1.5\,\%$ ($87.5\pm4.3\,\%$) compared to the baseline with $75\pm11\,\%$ ($61\pm 13\,\%$) for additional measurements from randomly selected configurations. The learning curves for states generated with the classical simulator and on a real quantum device are shown in Figures \ref{fig:ALLearningCurve} and \ref{fig:Example_GHZphase_IBMquantum}. In both cases it can be seen that adding samples which contain more information on the phase structure can confuse the learner at first place (see epochs $1000$ to $2000$ in both figures), but when enough information on the phase is added, the reconstruction improves again. For the state generated with the classical simulator (see Fig. \ref{fig:ALLearningCurve}) this happens by adding samples from the $xzzxx$ and $xyyyx$ configurations. For the states generated on the real devices five additional measurement configurations are needed.\\

For the state with all spins pointing upwards (all qubits having value one), the reference basis containing the most valuable information about the state is the $zzzzz$ basis. When using AL, this configuration is selected and the results are extremely good already in the first cycle of AL, with a fidelity of $f^{1/N}=99\,\%$ for simulator and real device. The QuCumber reconstruction without AL coincidentally uses the $zzzzz$ reference basis by default and hence coincidentally the perfect reference basis for the reconstruction of this state. Therefore, no difference between AL compared to the baseline can be observed.

For the state with all spins pointing in $x$ direction, the reconstruction results are improved as well. Here, the $xxxxx$ basis is chosen and the fidelity is increased by $14\,\%$ to $f^{1/N}=98.841\pm 0.022\,\%$ for simulated states ($10\,\%$ to $f^{1/N}=85.28\pm 0.40\,\%$ for real quantum devices) compared to a theoretically ideal state. Furthermore, the variance is decreased by a factor of up to $10$. In contrast to the GHZ state, where the improvement is achieved by increasing the fidelity step by step with every query, for this system the underlying reason for the improvement is the rotation of the reference configuration: When the $x$-spins state is rotated from the $zz\dots z$ configuration to the $xx\dots x$ configuration, the measurement distribution changes from equally distributed peaks for all outcomes to a peaked distribution at $00\dots 0$. Similarly to the state with all spins in $z$ direction, it is relatively easy for the RBMs to learn this distribution. We would like to point out that the reconstruction fidelities of the real quantum states are also limited by preparation errors, which becomes much more prominent for the $x$-spin states in contrast to the $z$-spin states, since a rotation from the $z$ to the $x$ direction comes with additional preparation errors. This is also in agreement with the large difference of reconstruction errors for simulated and real $x$-spin states.

\begin{table}[]
\footnotesize{
\begin{tabular}{l|ccc|ccc}
           & \multicolumn{3}{c|}{cl. quant. simulator} & \multicolumn{3}{c}{quant. device} \\
           & $N_{\mathrm{tot}}$ & $N_{\mathrm{queries}}$         & $N_{\mathrm{conf}}$      & $N_{\mathrm{tot}}$     & $N_{\mathrm{queries}}$   & $N_{\mathrm{conf}}$       \\\hline
GHZ        &    $62$            &    $2$               &     $3$    &    $102$          &   $2$  &     $3$                 \\
$\mathrm{GHZ}_{\varphi}$        &    $107$            &    $7$               &     $3$    &    $450$          &   $25$  &     $6$                 \\
$x$-spins &    $2$            &   $0$                   &     $1$ &     $2$         &   $0$           &     $1$        \\
$z$-spins   &   $4$             &     $0$ &     $1$                &        $4$      &    $0$   &     $1$  
\end{tabular}
}
\caption{Number of samples $N_{\mathrm{tot}}$, number of queries $N_{\mathrm{queries}}$ selected by the active learner and total number of configurations $N_{\mathrm{conf}}$ at the end of the learning for the reconstruction of the quantum states generated on IBM's classical quantum simulators and quantum devices. Here, the number of samples is $N_{\mathrm{per\,query}}=1$ except for the $\mathrm{GHZ}_{\varphi}$ state on the quantum device, where $N_{\mathrm{per\,query}}=10$. The reference configuration selected by the AL is $zz\dots z$ for $z$-spins, GHZ and $\mathrm{GHZ}_{\varphi}$ states, and $xx\dots x$ for the $x$-spins state. All states are defined in Eqs.~\eqref{eq:State1} to \eqref{eq:State3}.}
\label{tab:Samples_IBM}
\end{table}

\subsection{Tomography results for DMRG states \label{sec:DMRG}}


To investigate the AL reconstruction of many-body quantum states we can use the matrix product state framework for representing quantum states, such as for example ground states of the XXZ and the KCS models, and sampling in different basis configurations.
In this section, the AL results for these states with $5$ to $19$ qubits are presented in Figs.~\ref{fig:Heisenberg_summary} to \ref{fig:LGT_h=1}, with the number of samples and configurations used for the reconstruction from Tab.~\ref{tab:Samples_DMRG}. For all states a committee of four RBMs was used.

\subsubsection{Reconstruction of XXZ and Heisenberg States}
\begin{figure}[b]
	\centering
   \includegraphics[width=0.45\textwidth]{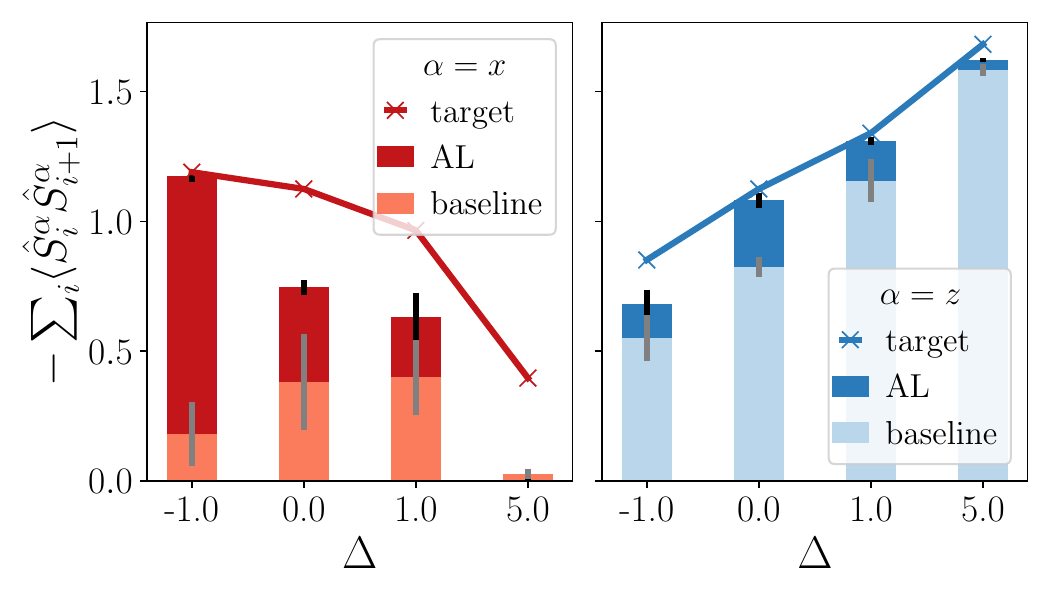}
	\caption[]{Reconstruction results for the ground states of the XXZ Hamiltonian \eqref{eq:Heisenberg}.}
	\label{fig:Heisenberg_summary}
\end{figure}

In figure \ref{fig:Heisenberg_summary} the reconstruction results for the ground states of the XXZ model with $8$ spins and different $\Delta=-1,0,1,5$ are presented. Here, we focus on the reconstruction of the correlator $\langle \hat{S}_i^\alpha \hat{S}_{i+1}^{\alpha}\rangle$ ($\alpha \in\{x,y,z\}$) for nearest neighbors and sum over all sites $i=1,\dots , L-1$. The learning is stopped when the reconstructed correlator $\vert \langle \hat{S}_i^\alpha \hat{S}_{i+1}^{\alpha}\rangle \vert $ is at least 2/3 of the target value and the correct sign is obtained. It can be seen that for $\Delta=-1,0,1$ the AL-QST scheme performs significantly better than the baseline, and for $\Delta=5$ equally well. Furthermore, we emphasize that the RBM representation in general yields better reconstruction results for the amplitudes than for the phases of the considered states, which has consequences for the reconstruction of the XXZ states: Firstly, the RBM reconstruction is not able to capture the SU(2) invariance of the $\Delta=0$ Heisenberg state, as can be seen e.g. also in Ref. \cite{Torlai2018}. Secondly, a better reconstruction of quantities measured in the reference direction compared to the orthogonal directions is obtained for all values of $\Delta$, e.g. the reconstruction of the correlator $\langle \hat{S}_i^\alpha \hat{S}_{i+1}^{\alpha}\rangle$ for $\Delta=0$ with $zz\dots z$ being the reference configuration has a value closer to the actual value for $\alpha=z$ than for $\alpha=x,\,y$ since the values for the latter correlators are systematically underestimated by the RBM representation. A similar tendency can be observed for small $\Delta$ like $\Delta=-1(1)$, where the reference configurations are selected by the AL-QST scheme to be the $xx\dots x$ ($zz\dots z$) configurations. Details on the AL-QST scheme can be found in Tab.~\ref{tab:Samples_DMRG}.

\begin{figure}[t]
	\centering
   \includegraphics[width=0.45\textwidth]{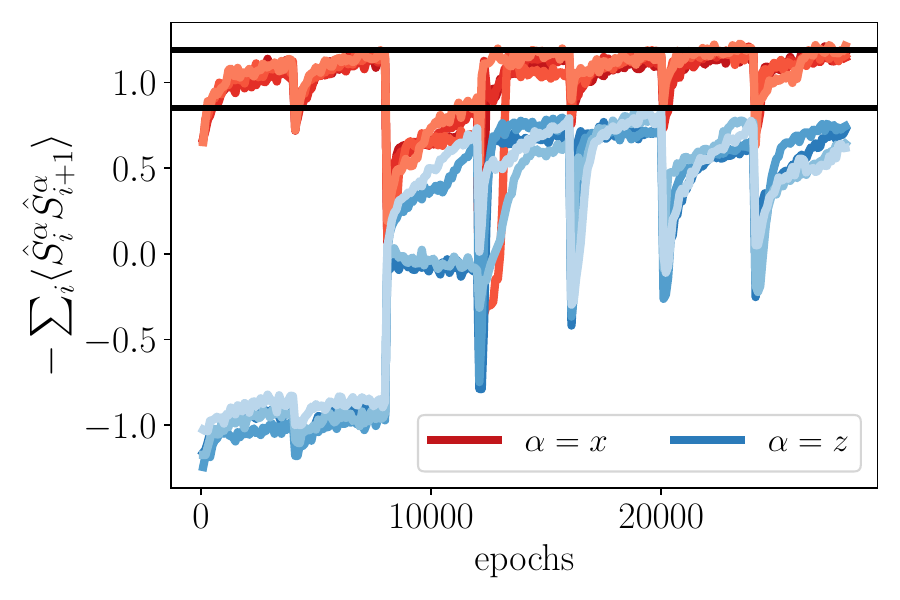}
	\caption[]{Learning curve for the ground state of the XXZ Hamiltonian \eqref{eq:Heisenberg} with $\Delta=-1$. The values for the target states are represented by the black lines.}
	\label{fig:Heisenberg_Delta=-1}
\end{figure}

The reconstruction curve for for $\Delta=-1$ is shown in Fig. \ref{fig:Heisenberg_Delta=-1}. This state has a small polarization in $z$ direction compared to $x$ and $y$ directions. Our AL-QST scheme is able to capture this by selecting the $xx\dots x$ configuration as the reference configuration and not the $zz\dots z$ configuration. Consequently, as explained above, the reconstruction of the correlator in the reference direction $\langle \hat{S}_i^x \hat{S}_{i+1}^{x}\rangle$ is better than e.g.  $\langle \hat{S}_i^z \hat{S}_{i+1}^{z}\rangle$. $\langle \hat{S}_i^z \hat{S}_{i+1}^{z}\rangle$ even has a wrong sign after the first learning phase up to epoch $3000$. After that, the AL-QST scheme requests the reference (i.e. $xx\dots x$) configuration in the first step (see epochs $3000$ to $6000$), which does not yield a significant improvement of the reconstruction of $\langle \hat{S}_i^z \hat{S}_{i+1}^{z}\rangle$. In the next two phases up to epoch $1200$ the AL scheme as explained in section \ref{sec:Model} selects the orthogonal configurations $zz\dots z$ and $yy\dots y$ (see epochs $6000$ to $12000$, which yields strongly improved results of $\langle \hat{S}_i^z \hat{S}_{i+1}^{z}\rangle$. In the last three queries measurements from the $xx\dots x$, $xxzxxzzx$ and $xzyzzzzz$ configurations improve the reconstruction of the correlators in all directions up to the threshold values and the learning procedure is stopped.



\subsubsection{Kinetically constrained spin chain model}

\begin{figure}[t]
	\centering
   \includegraphics[width=0.45\textwidth]{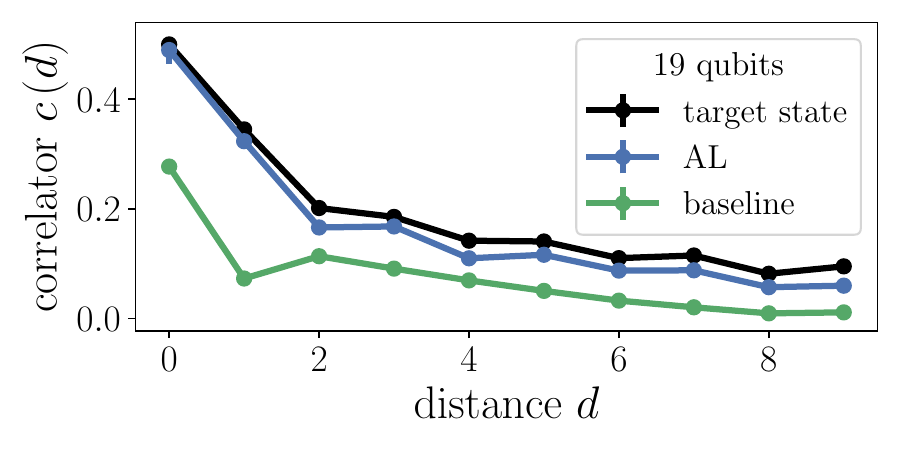}
   \includegraphics[width=0.45\textwidth]{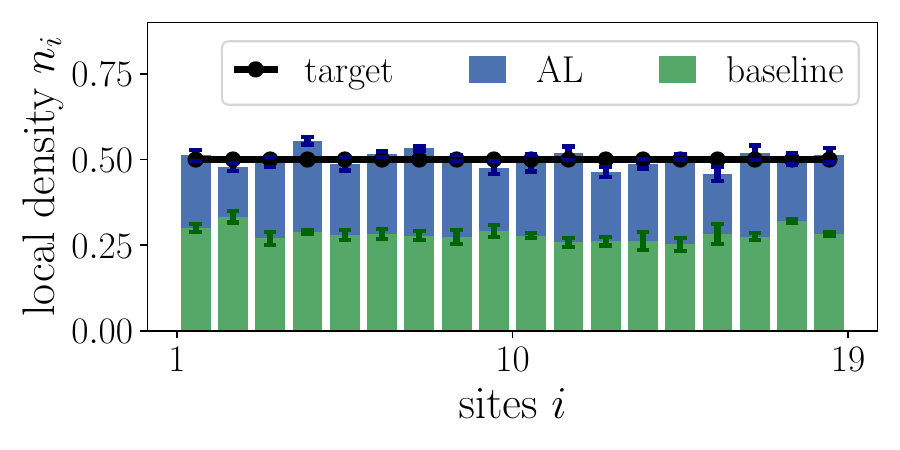}
	\caption[]{Correlator over distance (top) and density (bottom) from Eqs. \eqref{eq:GreensSpinDef} and \eqref{eq:LocalDensitySpinDef} for the KCS model state with $h/t=0$ and $19$ qubits for target state (black) and the reconstructed states using AL (blue) and the baseline (green).}
	\label{fig:LGT_h=0_2}
\end{figure}

\begin{figure}[t]
	\centering
   \includegraphics[width=0.45\textwidth]{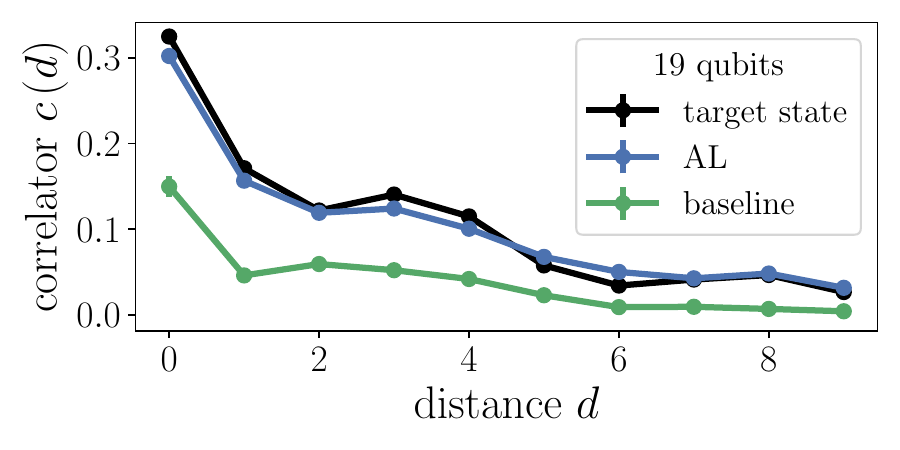}
   \includegraphics[width=0.45\textwidth]{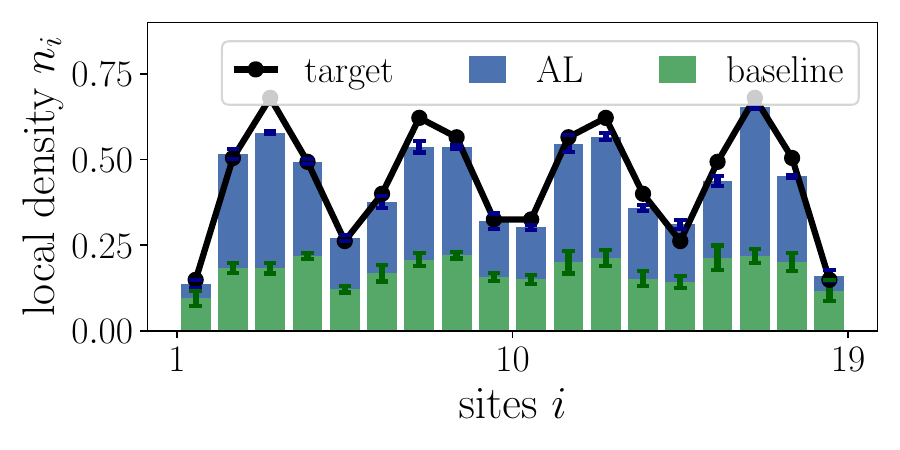}
	\caption[]{Correlator over distance (top) and density (bottom) from Eqs. \eqref{eq:GreensSpinDef} and \eqref{eq:LocalDensitySpinDef} for the KCS model state with $h/t=1$, $\mu/t=1$ and $19$ qubits for target state (black) and the reconstructed states using AL (blue) and the baseline (green).}
	\label{fig:LGT_h=1_2}
\end{figure}

\begin{figure}[t]
	\centering
\begin{minipage}[t]{0.23\textwidth}
   \includegraphics[width=1\textwidth]{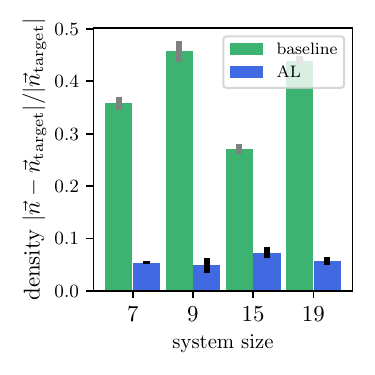}
\end{minipage}
\begin{minipage}[t]{0.23\textwidth}
   \includegraphics[width=1\textwidth]{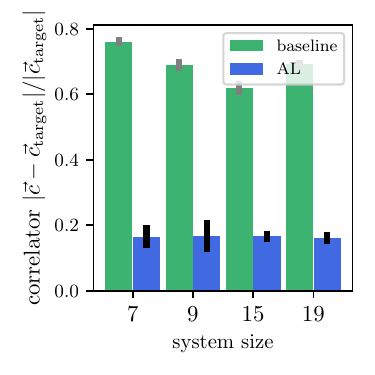}
\end{minipage}
	\caption[]{Relative difference of target and reconstructed densities (left) and correlators (right) from Eqs. \eqref{eq:LocalDensitySpinDef} and \eqref{eq:GreensSpinDef} of the KCS model with $h/t=0$. Error bars correspond to the standard error of the mean.}
	\label{fig:LGT_h=0}
\end{figure}

\begin{figure}[t]
	\centering
\begin{minipage}[t]{0.23\textwidth}
   \includegraphics[width=1\textwidth]{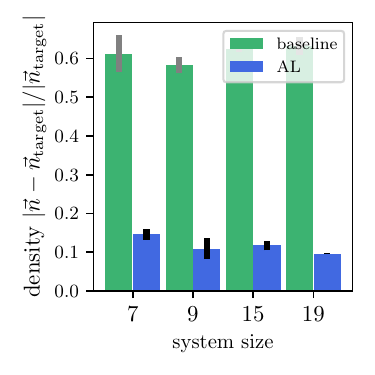}
\end{minipage}
\begin{minipage}[t]{0.23\textwidth}
   \includegraphics[width=1\textwidth]{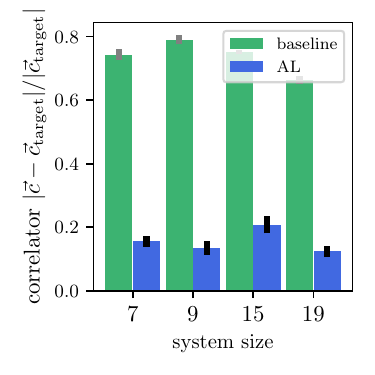}
\end{minipage}
	\caption[]{Relative difference of target and reconstructed densities (left) and correlators (right) from Eqs. \eqref{eq:LocalDensitySpinDef} and \eqref{eq:GreensSpinDef} of the KCS model with $h/t=1$ and $\mu/t=1$. An exemplary learning curve for $19$ qubits is shown in the Appendix \ref{appendix:MPS}.}
	\label{fig:LGT_h=1}
\end{figure}

In Figs.~\ref{fig:LGT_h=0_2} to \ref{fig:LGT_h=1} the tomography results for the kinetically constrained spin chain with $t=1$ and $h/t=0$, $\mu/t=0$ or respectively $h/t=1$, $\mu/t=1$ are summarized. For these states no full state vectors are available and hence the fidelity cannot be used to evaluate the quality of the reconstruction. Instead, we calculate the density and the correlator from Eqs. \eqref{eq:LocalDensitySpinDef} and \eqref{eq:GreensSpinDef} of the reconstructed states as defined in Sec.~\ref{sec:States} and compare them to the values for the target states. We have used the stopping conditions $\frac{\vert \bm{n}-\bm{n}_{\mathrm{target}}\vert}{\vert \bm{n}_{\mathrm{target}}\vert}\leq \tilde{n}_{\mathrm{stop}}=0.2$ and $\frac{\vert \bm{c}-\bm{c}_{\mathrm{target}}\vert}{\vert \bm{c}_{\mathrm{target}}\vert}\leq \tilde{c}_{\mathrm{stop}}=0.2$.

The target and reconstructed density $\bm{n}_{\mathrm{target}}$ and $\bm{n}$ have $N-1$ entries (for each possible domain wall between site $i$ and $i+1$). The spatially resolved domain wall densities for the kinetically constrained spin chain ground states with $h/t=0$, $\mu/t=0$ and $h/t=1$, $\mu/t=1$ are shown in Figs.~\ref{fig:LGT_h=0_2} and \ref{fig:LGT_h=1_2} (bottom) for a system with $19$ qubits. For $h/t=0$ and $\mu/t=0$ the local target density is $n_i=0.5$. For $h/t=1$ and $\mu/t=1$, the conserved total density is equal to $n^{\rm tot} = 8 / 18$ and one can observe Friedel oscillations, where the oscillations are proportional to $k_f = \pi n^{\rm tot}$ \cite{Borla2020}. For both target KCS model ground states the reconstructed states have a domain wall density which agrees with the target density in terms of magnitude and general characteristics (i.e. oscillations) when using AL. 

In contrast, the baseline without AL results in underestimated local densities $n_i$, which can be understood by the fact that the densities in Eq. \eqref{eq:LocalDensitySpinDef} are defined using $\hat{S}_i^x$ operators, which cannot be efficiently captured by the default $z$ reference configuration of the baseline RBMs. The AL learning scheme naturally overcomes this problem since it chooses the $xx\dots x$ configuration as the reference configuration. This selection of the reference configuration is only done based on the amplitudes and phases of the reconstructed wave functions at step 2 of the AL scheme. \\

A similar tendency can be observed for the target and reconstructed correlators $ \bm{c}_{\mathrm{target}}$ and $\bm{c}$. They have $\lfloor \frac{N}{2}\rfloor$ entries since we calculate the correlator over distance $d$ for the site at the middle of the chain. Also here the baseline reconstruction without AL yields values of $c(d)$ with a much smaller magnitude than for the target state for all distances $d$ and both parameter sets ($h/t=0$, $\mu/t=0$ and $h/t=1$, $\mu/t=1$) for the same reason as for the densities. In contrast, when using active learning the reconstructed correlator values are of the same magnitude as for the target state and even follow local features (see bending of the curve in Fig.~\ref{fig:LGT_h=1_2}). Moreover, the power-law for $h/t=0$ and exponential decay for $h/t=1$ can be reconstructed when using AL, but not for the baseline (see Appendix~\ref{appendix:MPS}).\\

To conclude, for the KCS model with $19$ qubits the reconstruction is improved drastically when using AL compared to the baseline scenario. This conclusion can also be drawn from the reconstruction of the KCS model states for other system sizes shown in Figs.~\ref{fig:LGT_h=0} and \ref{fig:LGT_h=1}. 
 For $h/t=0$ the change of reference basis yields a reduction of the density differences between target and reconstructed state by a factor $5$ for AL, from a relative value of around $44\,\%$ to around $6\,\%$ when averaging over all system sizes. The difference of correlators is decreased from $69\,\%$ to about $17\,\%$.  
For $h/t=1$ the difference in densities is decreased from an absolute value of around $61\,\%$ to less than $12\,\%$. The difference of target and reconstructed correlators is decreased to around $16\,\%$ from $74\,\%$. For both values of $h$ and all system sizes the active learner selected the $xx\dots x$ configuration as reference basis. This is different from the choice for the usual RBM procedure implemented in QuCumber which always uses the $zz\dots z$ configuration as reference basis. For almost all sizes the stopping condition was reached within the first learning cycle in step 3 (except for the $15$ qubit state ($h/t=0$), where another measurement in the reference basis was requested). Hence, this change of the reference frame by applying active learning already improves the results extremely, even without a need for the further steps 4 and 5.

\section{Summary and Outlook \label{sec:Summary}}

\begin{table}[t]
\footnotesize 
{
\begin{tabular}{p{2 cm}|p{1.2 cm}p{1.2 cm}p{1.2 cm}p{1.2 cm}}
\textbf{XXZ}                & $\Delta=-1$ & $\Delta=0$ & $\Delta=1$ & $\Delta=5$ \\\hline
reference        &    $xx\dots x$      &       $zz\dots z$  &  $zz\dots z$        &    $zz\dots z$      \\
$N_{\mathrm{tot}}$        &    $2000$      &       $2730$   &  $1750$        &    $1550$       \\
$N_{\mathrm{queries}}$  & $7$         &    $16$      & $6$         &  $30$  \\  
$N_{\mathrm{per\, query}}$  & $50$         &    $10$      & $50$         &  $50$  \\  
$N_{\mathrm{conf}}$  & $5$         &    $8$      & $4$         &  $1$ \\
\end{tabular}
\newline
\vspace*{0.2 cm}
\newline
\begin{tabular}{p{2 cm}|p{1.2 cm}p{1.2 cm}p{1.2 cm}p{1.2 cm}}
\textbf{KCS} ($\frac{h}{t}=0$)                & $7$ qubits & $9$ qubits & $15$ qub. & $19$ qub. \\\hline
reference        &    $xx\dots x$      &       $xx\dots x$  &  $xx\dots x$        &    $xx\dots x$      \\
$N_{\mathrm{tot}}$         &    $200$      &       $200$   &  $202$        &    $500$       \\
$N_{\mathrm{queries}}$  & $0$         &    $0$      & $1$  &  $0$   \\ 
$N_{\mathrm{per\, query}}$  & $-$         &    $-$      & $2$         &  $-$  \\ 
$N_{\mathrm{conf}}$  & $1$         &    $1$      & $1$         &  $1$         
\end{tabular}
\newline
\vspace*{0.2 cm}
\newline
\begin{tabular}{p{2 cm}|p{1.2 cm}p{1.2 cm}p{1.2 cm}p{1.2 cm}}
\textbf{KCS} ($\frac{h}{t}=1$)                & $7$ qubits & $9$ qubits & $15$ qub. & $19$ qub. \\\hline
reference   &    $xx\dots x$      &       $xx\dots x$  &  $xx\dots x$        &    $xx\dots x$      \\
$N_{\mathrm{tot}}$         &    $200$      &       $200$   &  $400$        &    $500$       \\
$N_{\mathrm{queries}}$  & $0$         &    $0$      & $0$  &  $0$   \\  
$N_{\mathrm{conf}}$  & $1$         &    $1$      & $1$         &  $1$         
\end{tabular}
}
\caption{Reference configuration selected by the AL, number of samples $N_{\mathrm{tot}}$, number of queries $N_{\mathrm{queries}}$, number of samples per query $N_{\mathrm{per\,query}}$ and configurations $N_{\mathrm{conf}}$ selected by the active learner for the reconstruction of the DMRG states: For the XXZ model and kinetically constrained spin (KCS) model ground states.}
\label{tab:Samples_DMRG}
\end{table}

In this work, we propose and implement an active learning scheme for adaptive quantum state tomography.
The active learning scheme uses the information available in the already measured data to propose the basis configuration for the next measurement with the most possible information gain. Inspired from the \textit{query-by-committee strategy} the information gain is calculated by taking the variance of reconstruction outcomes for different members of the committee into account.
We show that for a given number of measurements, our scheme provides a significant improvement in the reconstructed quantum state compared to a random choice of basis configurations. Our scheme brings the advantage that the information content of new measurement configurations is inferred from the variance of neural network quantum state representations based on previous measurements of the target state, and not from new measurements of the state under consideration, i.e. it allows to reduce the experimental effort. However, our implementation relies on the calculation of the variance for the set of potential next measurement configurations, scaling exponentially with the system size. This could be overcome by more advanced methods of exploring the space of measurement configurations, e.g. reinforcement learning, which will be considered in future work. Hereby, we imagine a reinforcement agent, trained to navigate in the space of all potential measurement configurations very efficiently, that selects possible candidates for next measurement configurations before applying our AL scheme.

The active learning scheme is generally applicable to different quantum states and devices, such as trapped ions, neutral atoms in optical tweezers, and superconducting qubits, as shown here. With the increasing number of quantum devices, the need for an efficient way to characterize the realized quantum state arises. Applications range from the verification of quantum computing devices, e.g. testing how faithfully a given quantum state can be prepared, to probing exotic states of matter realized in (analog) quantum simulators, such as the recently realized quantum spin liquid states \cite{Semeghini2021,Satzinger2021}, where measurements in different bases are necessary to characterize the quantum state. 

Apart from the implementation of our protocol in an interactive experimental feedback loop, possible directions for future work include more advanced schemes, e.g. more possible reference bases in the first step. Our active learning scheme can furthermore be generalized to state representations other than the restricted Boltzmann machines considered here, such as variational autoencoders \cite{Rocchetto2018}, recurrent \cite{Morawetz2021} and convolutional neural networks \cite{Schmale2021}, generative adversarial networks \cite{Ahmed2021}, and transformer architectures \cite{Cha2021}.

Another exciting future direction is the combination of the active learning scheme introduced here with the recently proposed classical shadows \cite{Huang2020} by extending the possible active learning actions to different unitary gates, potentially involving two or more qubits.  

\vspace{0.1cm}
\emph{Code availability.--} The Code for the AL algorithm is available at \url{https://github.com/HannahLange/Active-Learning}.

\emph{Acknowledgements.--}
We thank Tizian Blatz, Juan Felipe Carrasquilla, Roger Melko, Ejaaz Merali, Roger Luo, Anna Dawid, Ehsan Khatami, Sam Mardazad, Sebastian Paeckel, Henning Schlömer, and Jeffrey Thompson for fruitful discussions. This research was funded by the Deutsche Forschungsgemeinschaft (DFG, German Research Foundation) under Germany's Excellence Strategy -- EXC-2111 -- 390814868, by the European Research Council (ERC) under the European Union’s Horizon 2020 research and innovation programm (Grant Agreement no 948141) — ERC Starting Grant SimUcQuam and by the NSF through a grant for the Institute for Theoretical Atomic, Molecular, and Optical Physics at Harvard University and the Smithsonian Astrophysical Observatory.

\bibliographystyle{quantum}
\bibliography{main.bib}

\clearpage

\appendix 
\section{Quantum state representation in terms of a restricted Boltzmann machine \label{appendix:RBM}}

Part of our active learning scheme is the training of a committee of restricted Boltzmann machines (see Fig.~\ref{fig:AL}). In this work, the implementation of RBMs within the open-source software python package QuCumber \cite{Qucumber2019} is used for the training. The package is designed to learn quantum many body wave functions from a set of projective measurements in different basis configurations by representing the reconstructed state by RBMs. In this section, we will present how to represent quantum states in terms of RBMs and more details to the RBM training.\\

Firstly, a quantum state $\ket{\Psi}$ with only positive coefficients $\Psi(\bm{x}) = \langle\bm{x} \vert \Psi \rangle \geq 0$ will be considered, where 
$\{\ket{\bm{x}^b}\}$ with $\{\ket{\bm{x}^b}\} = \ket{x_1^b, \dots , x_N^b}$ is a reference basis $b$ for the Hilbert space of $N$ quantum degrees of freedom (i.e. for two qubits it consists of the states $\ket{00}$, $\ket{01}$, $\ket{10}$ and $\ket{11}$). In this case, the representation of a quantum state in terms of a RBM is straight forward: For an infinite number of measurements in a reference basis, i.e. in the $b = z$ basis, the measurements adhere to Born's rule and the probability of finding a measurement result $\bm{x}$ is 
$$
q(\bm{x})=\vert \Psi (\bm{x})\vert^2.$$

QuCumber creates a RBM network with a probability distribution given by the Boltzmann distribution
\begin{align}
    p_\lambda (\bm{x}, \bm{h})=\mathrm{exp}\left(\sum_{ij}W_{ij}h_ix_j+\sum_jb_jx_j+\sum_ic_i h_i \right)
\end{align}
and (by summing over the hidden nodes) the distribution over the visible nodes
\begin{align}
p_\lambda (\bm{x}) =&\frac{1}{Z}\prod_{j=1}^{V}\mathrm{exp}\left( b_j x_j\right) \\
&\quad \times \prod_{i=1}^{H} \left\{ 1 + \mathrm{exp}\left( \sum_j^V W_{ij} x_j +c_i \right)\right\},
\label{eq:RBM_prob2}
\end{align} 
where $\bm{v}$ and $\bm{h}$ are the visible and hidden nodes of the RBM and $W_{ij}$ the weights between visible node $i$ and hidden node $j$ \cite{Goodfellow} as shown in Fig. \ref{fig:RBM}. For a finite data set $\mathcal{D}= (\bm{x}_1, \bm{x}_2, \dots)$ it trains the RBM
such that the Kullback-Leibler divergence (Eq.~\eqref{eq:KL})
is minimized. The minimization is performed by gradient descent, which involves the calculation of expectation values with respect to the distributions of the data and the model. To calculate the expectation value over the model distribution one usually uses Markov Chain Monte Carlo sampling. Due to the restricted nature of RBMs hidden and visible units are conditional independent and hence the conditional probabilities factorize. Consequently, it is possible to calculate the conditional distributions of all visible / hidden nodes in parallel by taking $\bm{h}_{t+1} \propto p(\bm{h}\vert \bm{v}_t)$ and $\bm{v}_{t+1} \propto p(\bm{v}\vert \bm{h}_{t+1})$, where $t$ measures the number of steps in the Monte Carlo chain (block Gibbs sampling). For large $t\to \infty$ it is guaranteed to converge \cite{Mehta2019}. A slight modification which simplifies the training process is called contrastive divergence. Hereby, only $k$ iterations of Gibbs sampling are performed (contrastive divergence steps). For more details on the training procedure we refer the reader to Ref. \cite{Torlai2018}.\\

For positive wave functions the RBM representation can be defined as
$$
\psi_\lambda(\bm{x}) = \sqrt{\frac{p_\lambda(\bm{x})}{Z_\lambda}},
$$
with a normalization constant $Z_\lambda$ \cite{Qucumber2019}. 

For more general wave functions with complex-valued coefficients like
\begin{align}
\ket{\Psi} = \sum_{x_1, \dots x_N} \Phi_{x_1, \dots, x_N} \mathrm{e}^{i\varphi_{x_1, \dots, x_N}}\ket{x_1, \dots, x_N}
\label{eq:complex_psi}
\end{align}
the probability distribution underlying the outcomes of projective
measurements in the reference basis does not contain all possible information about the unknown quantum state, because the information about the phase is lost when only considering the underlying probability distribution of measurements in one basis $q(\bm{x})=\vert \Psi (\bm{x})\vert^2$. In this case QuCumber represents the quantum state as defined in Eq.~\eqref{eq:RBMstate}
and trains two RBMs with parameters $\lambda$ and $\mu$ separately. The RBM with parameters $\lambda$ models the amplitude of the RBM wave function, the second RBM with parameters $\mu$ the phase $\theta_\mu(\bm{x}) = \mathrm{log}\, p_\mu(\bm{x})$. Firstly $\lambda$ is optimized such that $\vert \Psi (\bm{x}^b)\vert ^2 = \vert \psi_{\lambda, \mu} (\bm{x}^b)\vert ^2 $ for measurements $\bm{x}^b$ in the reference basis $b$, which is set to the $z$-basis by default. Secondly, measurements in other basis configurations are considered to determine the phase $\theta_\mu$ by training another RBM \cite{Torlai2018}.\\

\begin{figure}[t!]
	\centering
  \includegraphics[width=0.4\textwidth]{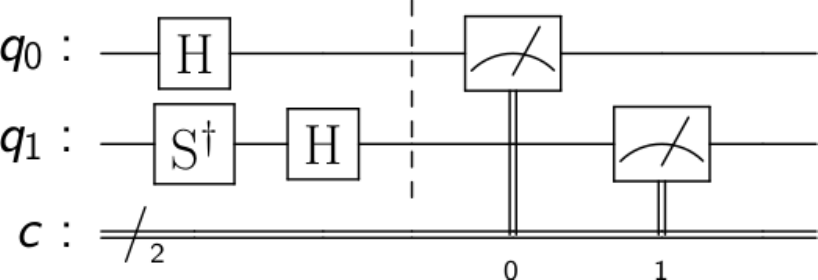}
	\caption[Measurement of a quantum circuit in the $xy$ basis configuration]{Measurement of a quantum system with two qubits ($q_0$ and $q_1$) in the $xy$ basis configuration. Therefore $q_0$ is rotated from the $z$ to the $x$ axis by applying the Hadamard gate, and $q_1$ to the $y$ axis by application of a combination of $S$ and Hadamard gates (see equations \eqref{eq:Hadamard} and \eqref{eq:K}). Figure generated with Qiskit \cite{Qiskit2010}.}
	\label{fig:Configs}
\end{figure}

The information about the phase is extracted via the rotation of single qubits within a quantum circuit into another basis. QuCumber uses by default the $z$ basis as reference basis and rotates to the $x$ and $y$ bases to extract phase information. The rotation of a single qubit to the $x$ basis is achieved by applying the Hadamard gate 
\begin{align}
    H = \frac{1}{\sqrt{2}}\left( \begin{array}{rr}
1 & 1  \\ 
1 & -1 \\
\end{array}\right),
\label{eq:Hadamard}
\end{align}
the rotation to $y$ by applying a combination of the Hadamard gate and the S-adjoint gate,
\begin{align}
    K = \frac{1}{\sqrt{2}}\left( \begin{array}{rr}
1 & 1 \\ 
i & -i \\
\end{array}\right).
\label{eq:K}
\end{align}

The rotations are performed by using the pre-defined quantum operations of Qiskit as shown for an exemplary rotation from $zz$ to $xy$ in figure \ref{fig:Configs}.
For the DMRG states, 
we use
\begin{align}
    R_x = \frac{1}{\sqrt{2}}\left( \begin{array}{rr}
i & -i \\ 
1 & 1 \\
\end{array}\right)
\label{eq:R_x}
\end{align}
and 
\begin{align}
    R_y = \frac{1}{\sqrt{2}}\left( \begin{array}{rr}
1 & -i \\ 
-i & 1 \\
\end{array}\right)
\label{eq:R_y}
\end{align}
as local basis rotation matrices.

The representation of a complex wave function in terms of the RBMs as explained above is implemented with the QuCumber package by using the \texttt{ComplexWaveFunction} method.

\section{Details of active learning QST}


When using the AL procedure as described in section \ref{sec:Model}, within each cycle (step 3), $4$ RBMs are used if not stated otherwise, with
$1000$ epochs, a learning rate of $l = 0.07$  and contrastive divergence steps $k = 100$.   

\subsection{IBM quantum states \label{appendix:IBM}}

\begin{figure}[t]
	\centering
\includegraphics[width=0.5\textwidth]{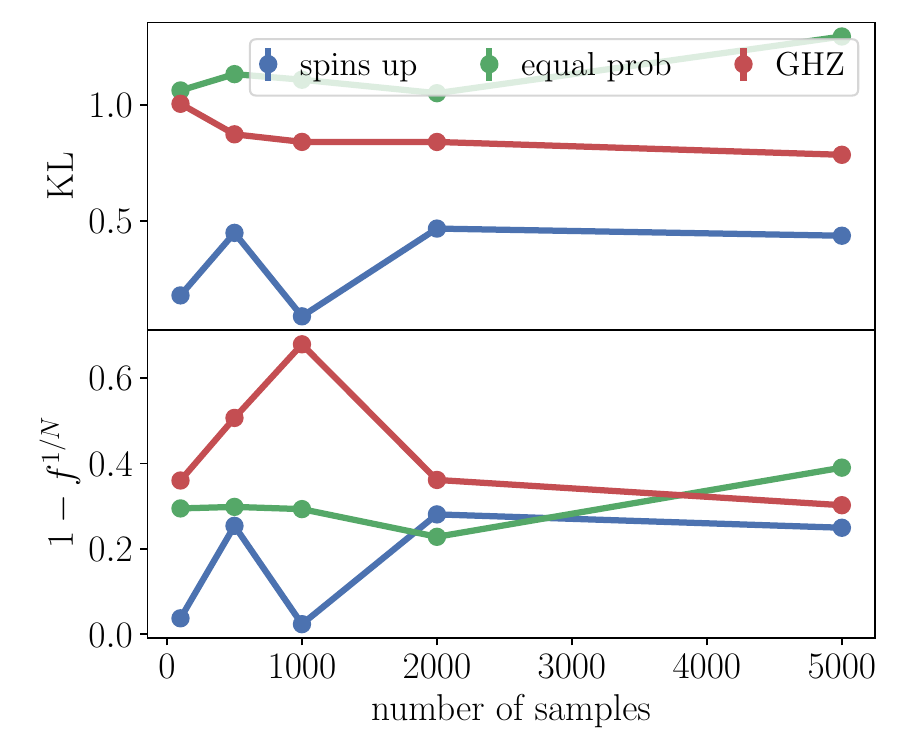}
	\caption[]{Reconstruction results for a pure RBM reconstruction of quantum states prepared on an IBM device ($5$ qubits) without AL. The number of samples is varied by keeping the number of configurations fixed ($6$ configurations).}
\label{fig:WithoutAL1}
\end{figure}
\begin{figure}[t]
	\centering
\includegraphics[width=0.5\textwidth]{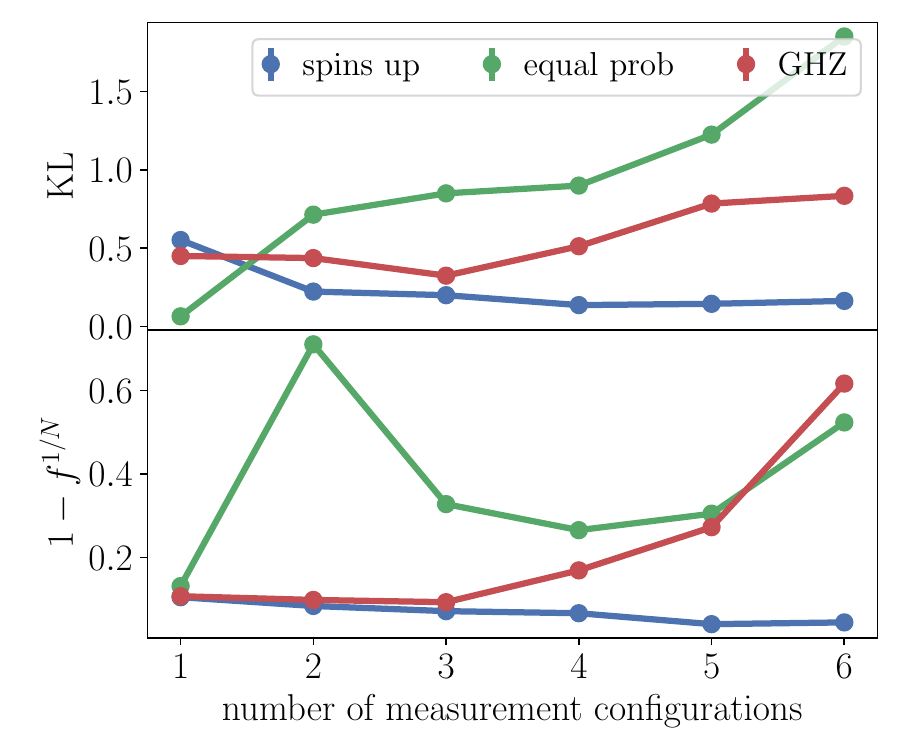}
	\caption[]{Reconstruction results for a pure RBM reconstruction of real quantum states ($5$ qubits) without AL. The number of configurations is varied by keeping the number of samples fixed ($2000$ samples).}
\label{fig:WithoutAL2}
\end{figure}

For the generation of states on a classical quantum simulator we use the Aer simulator, which is designed to  mimic the execution of an actual device \cite{IBM}. For real quantum states the devices \textit{ibmq bogota} and \textit{ibmq quito} were used, which both consist of $5$ superconducting qubits. 

In figures \ref{fig:WithoutAL1} and \ref{fig:WithoutAL2} the results for a pure RBM reconstruction of real quantum states ($5$ qubits) without AL are shown. In \ref{fig:WithoutAL1} the number of samples is varied by keeping the number of configurations fixed ($6$ configurations). In \ref{fig:WithoutAL2} the number of configurations is varied when fixing the number of samples ($2000$ samples). It can be seen that it is difficult and very time consuming to find the most efficient set of measurements and configurations by such scans. Furthermore, the number of samples and configurations with the best reconstruction varies from state to state. This makes our active learning scheme very appreciable since it chooses the number of samples and configurations on its own and only the number of samples at the beginning and per query have to be fixed by the user. As can be seen from section \ref{sec:States} equally good results can be obtained for all states when setting the number of samples per query to $1$ or $2$. Hence, the only free parameter is the number of samples at the beginning of the AL.  


\subsection{DMRG states \label{appendix:MPS}}

\begin{figure}[t]
	\centering
  \includegraphics[width=0.48\textwidth]{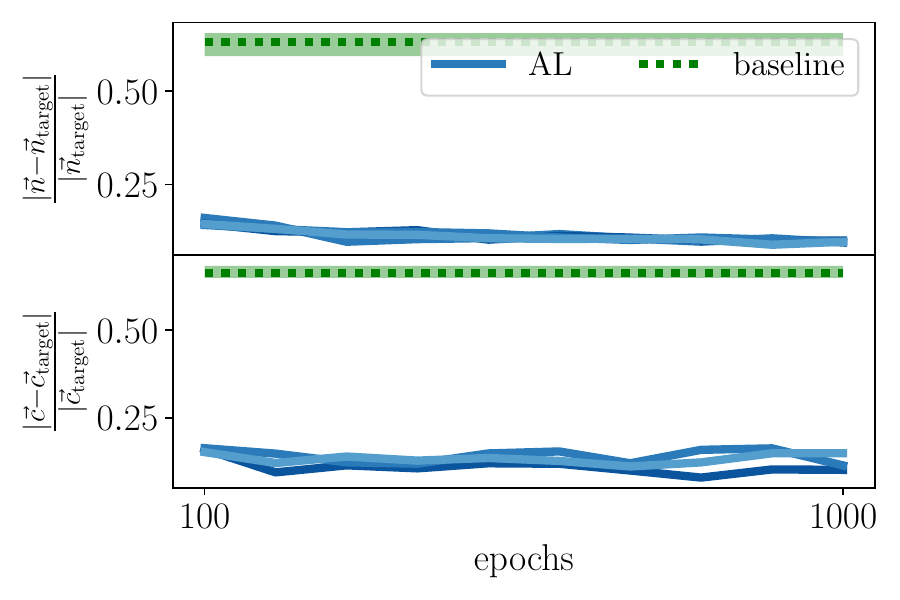}
	\caption[]{Exemplary learning curves for a LGT state with $19$ qubits and $h/t=1$.}
	\label{fig:Example4}
\end{figure}

\begin{figure}[t]
	\centering
   \includegraphics[width=0.45\textwidth]{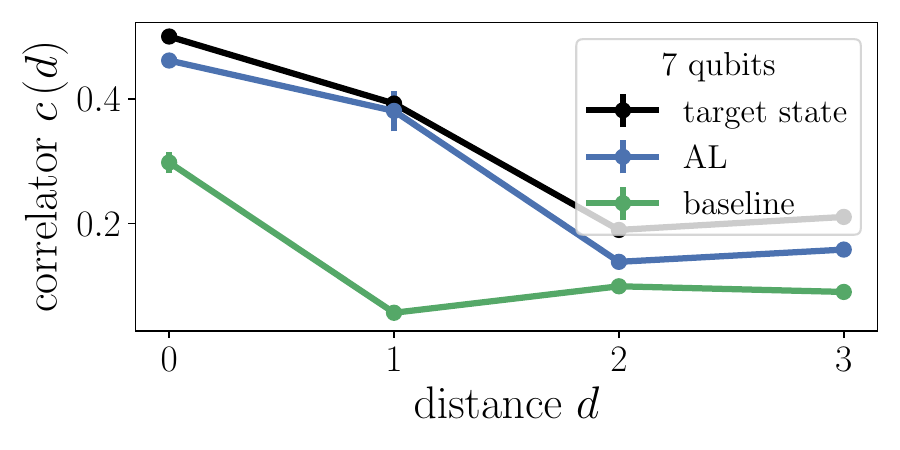}
   \includegraphics[width=0.45\textwidth]{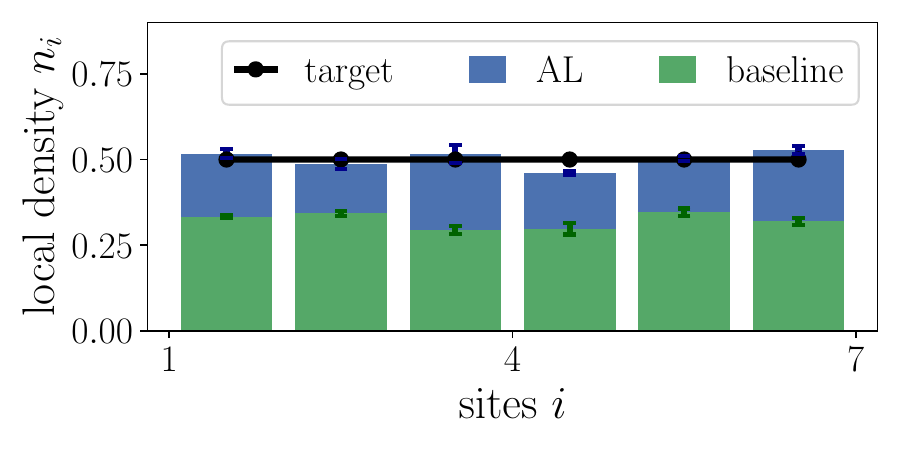}
	\caption[]{Correlator over distance (top) and density (bottom) for the KCS state with $h/t=0$ and $7$ qubits for target state (black) and the reconstructed states using AL (blue) and the baseline (green).}
	\label{fig:LGT_h=0_3}
\end{figure}

\begin{figure}[t]
	\centering
   \includegraphics[width=0.45\textwidth]{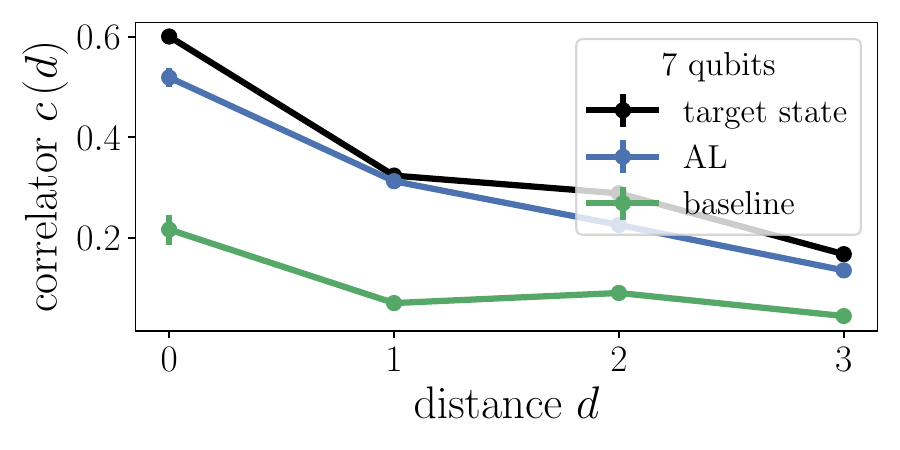}
   \includegraphics[width=0.45\textwidth]{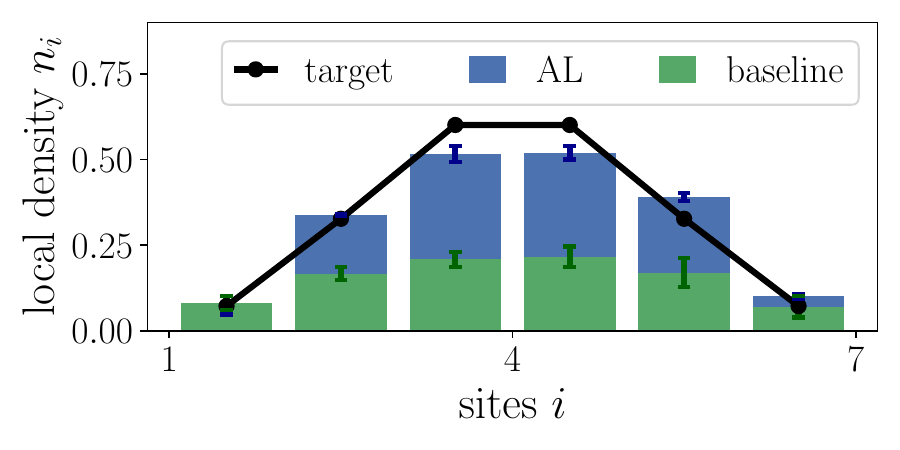}
	\caption[]{Correlator over distance (top) and density (bottom) for the KCS state with $h/t=1$, $\mu/t=1$ and $7$ qubits for target state (black) and the reconstructed states using AL (blue) and the baseline (green).}
	\label{fig:LGT_h=1_3}
\end{figure}


For lattice gauge model states we consider states with $\mu=10^{-7}\approx 0$, since the convergence is much faster than for $\mu=0$. 

In Fig.~\ref{fig:Example4} the learning curve for a lattice gauge model state with $19$ qubits and $h/t=1$ is shown. When using AL divergence, the $xxxxxxx$ configuration is chosen to be the reference basis. One can see that only this choice of reference basis leads to relatively good results for the density differences (around $15\,\%$, see upper part for epoch $0$) and $17\,\%$ correlator difference (bottom). The RBM training then reduces the density difference further to $\vert \bm{n}-\bm{n}_{\mathrm{target}}/\vert \bm{n}_{\mathrm{target}}\vert= 9.59\pm 0.02\,\%$ and $\vert \bm{c}-\bm{c}_{\mathrm{target}}/\vert \bm{c}_{\mathrm{target}}\vert= 12.4\pm 0.2\,\%$. \\

In Figs.~\ref{fig:LGT_h=0_3} and \ref{fig:LGT_h=1_3} the local correlators over distance $d$ (top) and densities over the system size are shown for KCS states with $7$ qubits and $h/t=0$, $\mu/t=0$ and $h/t=1$, $\mu/t=1$ respectively. Similarly to the results for $19$ qubits in Figs.~\ref{fig:LGT_h=0_2} and \ref{fig:LGT_h=1_2} one can see that the agreement with the target state is improved when using the AL reconstruction scheme. Also here, features like the curvature of the correlator can be reproduced for AL, but not for the baseline. For the parameters $h/t=1$ and $\mu/t=0$ the density increases to a maximum in the middle of the chain. Even though this behaviour can be reproduced by AL and baseline, the results for the baseline are of a around half of the magnitude of the target state. In contrast, the AL reconstruction yields densities with magnitudes much closer to the target state.

\begin{figure}[t]
	\centering
   \includegraphics[width=0.45\textwidth]{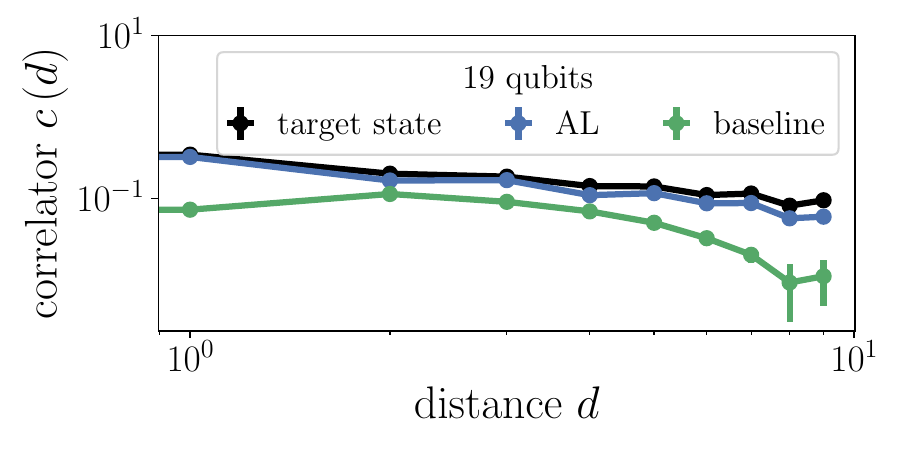}
	\caption[]{Correlator over distance for the KCS model state with $h/t=0$, $\mu/t=0$ and $19$ qubits for target state (black) and the reconstructed states using AL (blue) and the baseline (green) as in Fig.~\ref{fig:LGT_h=0_2} but with logarithmic scaling for $x$ and $y$ axes.}
	\label{fig:LGT_h=0_21}
\end{figure}

\begin{figure}[t]
	\centering
   \includegraphics[width=0.45\textwidth]{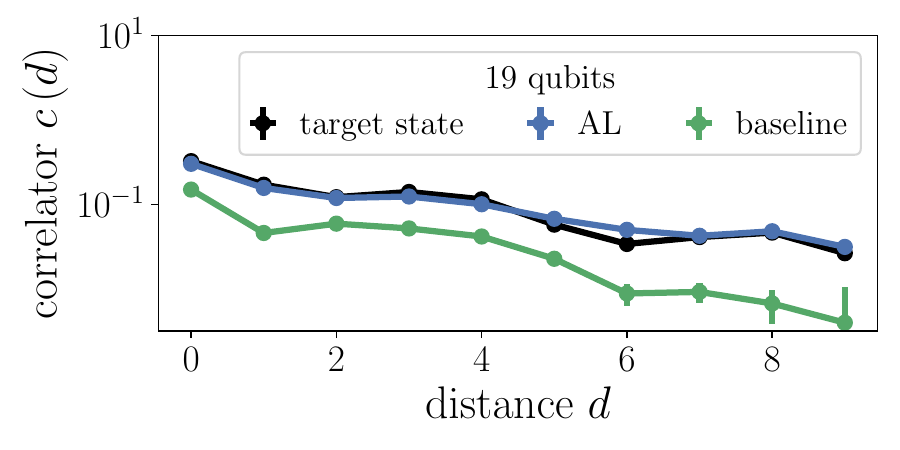}
	\caption[]{Correlator over distance for the KCS model state with $h/t=1$, $\mu/t=1$ and $19$ qubits for target state (black) and the reconstructed states using AL (blue) and the baseline (green) as in Fig.~\ref{fig:LGT_h=1_2} but with logarithmic scaling for the $y$ axis.}
	\label{fig:LGT_h=1_21}
\end{figure}

Moreover, when plotting the correlators for $h/t=0$ ($h/t=1$) with logarithmic scales for $x$ and $y$ axes ($y$ axis) as shown for a system with $19$ qubits in Fig.~\ref{fig:LGT_h=0_21} (Fig.~\ref{fig:LGT_h=1_21}) one can observe the expected power-law (exponential) decay for the target state. This power-law (exponential) decay is reconstructed when using AL, but not for the baseline.

\section{Kinetically constrained spin model and $\mathbb{Z}_2$ lattice gauge theory}
\label{ApdxKinConsSpn}
In this section we provide more background information on the kinetically constrained quantum spin model, see Eq.~\eqref{eq:LGT_Model_Spin_Def}, considered in the main text.

\subsection{Mapping to a $\mathbb{Z}_2$ lattice gauge theory}
The model in Eq.~\eqref{eq:LGT_Model_Spin_Def} can be mapped to a one-dimensional \Zt lattice gauge theory model, with U(1) matter \cite{Borla2020,Kebric2021}. To this end, domain walls in the spin model are mapped to hardcore bosons, which are coupled to \Zt gauge fields defined on the links between the lattice sites $i,j$. As explained below, one obtains the following equivalent Hamiltonian,
\begin{align}
    \H_{\mathbb{Z}_2} = & -t \sum_{\langle i, j \rangle} \l \ad_i \hat{\tau}^{z}_{\langle i, j \rangle} \a_j + \hc \r
    \notag \\
    &- h \sum_{\langle i, j \rangle}
    \hat{\tau}^{x}_{\langle i, j \rangle}
    + \mu \sum_{\langle i, j \rangle} \hat{n}_j.
    \label{eq:LGT_Model_Original_Def}
\end{align}
Here $\ad_j$ is the hardcore boson creation operator, defined on site $j$, and $\hat{n}_j = \ad_j \a_j$ is the local number operator. The \Zt gauge and electric fields are represented with the Pauli matrices, defined on the links between neighboring lattice sites, as $\hat{\tau}^{z}_{\langle i, i+1 \rangle}$ and $\hat{\tau}^{x}_{\langle i, i+1 \rangle}$ respectively. This model is appealing since it exhibits confinement of dynamical particles which is induced by any nonzero \Zt electrical field term $h \neq 0$ \cite{Borla2020,Kebric2021}.

The generator of the local \Zt gauge symmetry of Eq.~\eqref{eq:LGT_Model_Original_Def} can be written as \cite{Prosko2017}
\begin{equation}
    \hat{\mathcal{G}}_j = \hat{\tau}^{x}_{\langle i-1, i \rangle}
    (-1)^{\hat{n}_i} \hat{\tau}^{x}_{\langle i, i+1 \rangle}, \qquad [\H_{\mathbb{Z}_2},\hat{\mathcal{G}}_j]=0.
    \label{eq:GaussLaw}
\end{equation}
This leads to the \Zt Gauss law, requiring all states to be $+1$ eigenstates of $\hat{\mathcal{G}}_j$ (we assume no background charges):
\begin{equation}
    \hat{\mathcal{G}}_j \ket{\psi} = + \ket{\psi}.
    \label{eq:GaussLawConstr1}
\end{equation}

From here it is relatively straightforward to obtain back the constrained spin model Eq.~\eqref{eq:LGT_Model_Spin_Def} from the main text. One notices that the charge configuration is entirely determined by the \Zt electric fields due to the constraints imposed by the \Zt Gauss law. This allows to formulate the Hamiltonian entirely in terms of the gauge field, by identifying  the presence of a particle on a lattice site as anti-alignment of the \Zt electric field defined on the links connecting that site. This also means that the particles, or domain walls, are connected with the \Zt electric fields of the same orientation, which we interpret as strings and anti-strings which connect the \Zt charges.

Using the above interpretation it is straightforward to see that the first term in model \eqref{eq:LGT_Model_Spin_Def} corresponds to the kinetic term, where the number of domain walls is conserved. This ensures two things in the lattice gauge interpretation: the \Zt electric string remains attached to the hopping particle, and the total number of particles is conserved.
The second term $\propto h$ in Eq.~\eqref{eq:LGT_Model_Spin_Def} induces energy cost to strings and as a result confines the particle pairs into dimers. Finally, the Ising term is needed to control the number of domain walls and thus the filling of the chain in the 1D LGT interpretation. 

\subsection{The gauge-invariant equal-time Green's function}

The correlation function in Eq.~\eqref{eq:GreensSpinDef} can be mapped to the \Zt invariant equal-time Green's function defined as
\begin{equation}
    {\bm{c}(|i-j|) = \left \langle \ad_i \prod_{i \leq l < j} \hat{\tau}^{z}_{\langle l, l+1 \rangle} \a_j \right \rangle}.
\end{equation}
This is once again done by taking into account the constraint imposed by the Gauss law in Eq.~\eqref{eq:GaussLawConstr1}.
Both terms  $\propto \l 1 \pm 4\hat{S}^{x}_{l}\hat{S}^{x}_{l+1}\r, l \in \{i, j \}$ act as projectors to the states where there is a site with a particle to be annihilated and an empty site, where a particle can be created. The actual annihilation at site $j$ and creation of the particle at site $i$ comes from the application of $\prod_l \l 2 \hat{S}^{z}_l \r$ between the domain walls. Combining both terms, first applying the projectors and then performing the annihilation and creation of the particle, gives us the \Zt Green's function expressed entirely in terms of gauge fields.
Note that the order is slightly changed in the main text Eq.\eqref{eq:GreensSpinDef} by considering the spin anticommutation property.

Such correlation function probes the confinement of particle pairs in the LGT interpretation. It decays as a power law in the deconfined phase $h = 0$ and decays exponentially in the confined phase, which is the case for any non-zero value of the field \cite{Borla2020, Kebric2021}. On the other hand, the pair-pair correlation function decays as a power-law function also in the confined phase, which means that the effective dimers behave as Luttinger liquid \cite{Borla2020}.

\end{document}